\documentclass[letterpaper, 10 pt, conference]{ieeeconf}
\IEEEoverridecommandlockouts 

\usepackage{blindtext, graphicx}
\usepackage{amsmath}
\usepackage{amssymb}
\usepackage{epstopdf}
\usepackage{mathtools}
\usepackage{algorithm}
\usepackage[noend]{algpseudocode}
\usepackage{cite}
\usepackage{amssymb}
\usepackage{color}
\newtheorem{thm}{Theorem}
\newtheorem{lem}[thm]{Lemma}
\newtheorem{prop}[thm]{Proposition}

\newtheorem{def1}{Definition}
\newtheorem{exmp}{Example}

\newtheorem{rem}{Remark}

\newtheorem{problem}{Problem}
\newcommand{\ind}{\operatorname{ind}}
\newcommand{\im}{\operatorname{im}}

\makeatletter
\def\BState{\State\hskip-\ALG@thistlm}
\makeatother
\algnewcommand\algorithmicinput{\textbf{Input:}}
\algnewcommand\INPUT{\item[\algorithmicinput]}
\algnewcommand\algorithmicoutput{\textbf{Output:}}
\algnewcommand\OUTPUT{\item[\algorithmicoutput]}

\title{\LARGE \bf Control refinement for DAE systems: A behavioral approach via simulation relations}
\author{Fei Chen, \textit{Systems \& Control, TU/e}\\
Department of Electrical Engineering, Eindhoven University of Technology, the Netherlands}

\begin{document}
\maketitle
\thispagestyle{plain}
\begin{abstract}
The controller design of the so-called ``difference algebraic equation'' (DAE) systems that are frequently shown in industrial processes, tend to be challenging because of the combination of algebraic equations and high state dimensions. In this paper, we tackle this problem by developing control refinement approaches for DAE systems via the notions of (bi)simulation relations and approximate simulation relations from computer science. The quantified refinement accuracy is achieved by defining observation metrics over a general system framework named transition systems. We employ the behavioral theory to tackle dynamical systems and control problems in a more general framework. Due to the difficulty in dealing with a DAE system directly, we derive another system, which is behaviorally equivalent to the related DAE system and in standard state space form, to provide ease for further control refinement. Consequently, well-developed model reduction approaches can be applied to obtain an abstract simplified system, which can be rewritten into a DAE system again. Based on the (bi)simulation relations, approximate simulation relations and the initialization conditions, we show that for any given well-posed controller of the abstract model, we can always refine it to a controller for the concrete model such that the two systems have the same controlled output behavior or the distance between their output behavior is bounded. 
\end{abstract}


\section{Introduction}
Industrial processes tend to have models with huge complexity and state dimensions, and usually contain algebraic equations in addition to difference equations. These, so-called ``difference algebraic equations'' (DAE)~\cite{kunkel2006differential,dai1989singular}, are also common in some mechanical systems like cars and robots. Actually, the combination of algebraic equations and high state dimensions make numerical simulation and controller design of DAE systems challenging if not impossible. Hence, industry needs for methods to resolve the simulation and controller design problems posed by these complex DAE models.

For models solely composed of ``ordinary difference equations'' (ODE), the rapidly developing model reduction methods such as proper orthogonal decomposition (POD), balanced truncation, Hankel norm model reduction, etc,~\cite{antoulas2005approximation} can be applied to derive the reduced order models. These models can be used to provide ease in modelling, simulation and design. However, when dealing with complex DAE systems, these model reduction methods for ODE systems cannot be applied directly. There does exist some research regarding the model reduction approaches for DAE systems, but not that widely developed. For instance,~\cite{stykel2004gramian} proposes a gramian-based model reduction method. On the other hand, ~\cite{Xingangthesis} presents Hankel norm model reduction approaches based on system decompositions via the so-called Weierstrass canonical form.   

In industry, engineers usually regard DAE models as dynamical systems with some constraints and deal with them by writing the algebraic equations in explicit forms. By substituting the explicit expressions in the dynamical equations, the original models are recast as ODE systems and then controller strategies can be designed. For example, in~\cite{porru2015quality}, the author employs this method to tackle nonlinear DAE models representing industrial multicomponent distillation columns. However, in general, when we deal with complex DAE systems that show huge dimensions in the algebraic part, this method usually does not make sense due to the fact that the explicit expressions cannot be always found.   

\begin{figure}[!h]
\centering
\includegraphics[width=0.5\columnwidth]{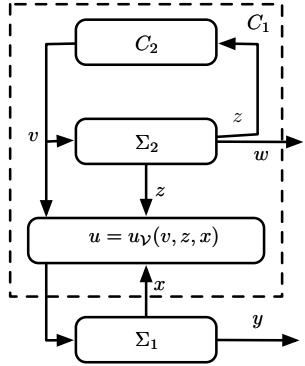}
\caption{Hierarchical control for ODE systems.}
\label{hie1}
\end{figure}

Therefore, in this paper, we tackle controller design problems of complex DAE models by developing control refinement approaches. Consider a complex DAE model and its reduced order model in DAE representation; control refinement means finding a general method to refine a well-posed controller for the reduced model to obtain another controller for the original model. Actually, it is hard to deal with DAE systems directly and in discrete-time, DAE systems show anti-causality~\cite{dai1989singular}. Therefore, the behavioral theory\cite{willems2013introduction, willems2007behavioral, willems1991paradigms, willems1998quadratic}, which makes a formal distinction between a system (its behavior) and its representations, is investigated to treat DAE systems and control problems in a more general framework. The notions of (bi)simulation relations and approximate simulation relations~\cite{metric, bridge} from computer science establish relationships between two systems and could be connected with the behavioral theory. For instance, in~\cite{tabuada}, the output behavior is connected with these notions. Inspired by these notions, we are interested in how to establish ``bridges'' between systems to benefit the further control refinement. In~\cite{van2004equivalence} and~\cite{megawati2015bisimulation}, the authors discuss the bisimulation equivalence of nondeterministic ODE and DAE systems, respectively. In addition, approximate (bi)simulations for constrained linear systems and nonlinear systems are proposed in~\cite{linearconstrain} and~\cite{nonlinear}, respectively. For the application of these relations in control problems, the hierarchical control framework for continuous-time ODE systems as shown in Figure \ref{hie1} is presented in~\cite{main}. This framework gives us a lot of insights to develop control refinement approaches for DAE systems. On the other hand, in~\cite{fainekos2007hierarchical}, the author uses these notions to  tackle the problem of synthesizing a hybrid controller based on a specification that is expressed as a temporal logic formula.    

In this paper, we deal with DAE systems within the behavioral framework and we are interested in how to develop exact and approximate control refinement approaches for DAE systems via the notions of (bi)simulation relations and approximate simulation relations.

%

The structure of this paper is as follows. We close this section with the mathematical notations used in this paper. Section \uppercase\expandafter{\romannumeral2} introduces the framework of behavioral approach and formulates our problems. In Section \uppercase\expandafter{\romannumeral3}, the properties of DAE systems and the notions of (bi)simulation relations, approximate simulation relations and simulation functions are presented. Section \uppercase\expandafter{\romannumeral4} is dedicated to the exact control refinement for DAE systems. In Section \uppercase\expandafter{\romannumeral5}, the hierarchical control for discrete-time ODE systems is presented and afterwards the approximate control refinement approach for DAE systems is developed. The last section closes with the concluding remarks and the future work.
\subsection*{Notation}
Following concepts will be used throughout this paper.
\begin{itemize}
\item $\mathbb{T}$ is the time with $\mathbb{T}:= \mathbb{N}_0$.
\item $d(x_1,x_2)$ is a distance function or a metric defined over two vectors in the same Euclidean space.
\item Unless stated otherwise, $u: \mathbb{T}\rightarrow \mathbb{R}^m$ represents a time dependent signal or sequence, which maps the time to some Euclidean spaces such that $u(t)\in \mathbb{R}^m$ with $t\in \mathbb{T}$.
\item $\|\cdot\|$ stands for the Euclidean norm for a vector with the triangle inequality $\|x+y\|\leq\|x\|+\|y\|$. The induced metric is defined as $d(x_1,x_2)=\|x_1-x_2\|$. The induced 2-norm of a matrix is denoted by $\|\cdot\|_2$. 
\item The supremum norm of a signal $u$ denoted by $u_{\max}$ is defined as $$u_{\max}=\max_{t\in \mathbb{T}}\|u(t)\|.$$
\item Given a metric space $X$, the $\varepsilon$-ball $B_\varepsilon(x)$ of radius $\varepsilon>0$ with center $x\in X$ is defined as $B_\varepsilon(x)=\{y\in X\mid \|x-y\| \leq \varepsilon\}.$ For a set $A\subset X$, $\mathcal{C}_\varepsilon(A)=\{x\in X\mid B_\varepsilon(x) \subseteq A\}$ is called the $\varepsilon$-contraction of $A$ and $\mathcal{E}_\varepsilon(A)=\{x\in X\mid B_\varepsilon(x) \cap A\neq \emptyset\}$ is called the $\varepsilon$-expansion of $A$.
\item For two sets $X_1$ and $X_2$ with the Cartesian product defined as $X_1\times X_2=\{(x_1,x_2)\mid x_1\in X_1, x_2\in X_2\}$. A relation $\mathcal{R}\subseteq X_1\times X_2$ is a subset of this Cartesian product that relates the elements $x_1\in X_1$ with the elements $x_2\in X_2$.
\end{itemize}

\section{Framework $\&$ Problem statement}
\noindent In the very beginning of this section, we talk about behavioral theory as it introduces a general framework to treat dynamical systems. This framework can be used later to define DAE systems in the behavioral point of view. Finally, the problem statement is formulated based on the developed behavioral framework. 

\subsection{Behavioral theory}
\begin{def1}
\textit{~\cite{willems2013introduction} A dynamical system $\Sigma$ is defined as a triple
\begin{equation}\nonumber
\Sigma=(\mathbb{T},\mathbb{W},\mathfrak{B})
\end{equation}
with $\mathbb{T}$ a subset of $\mathbb{R}$ or $\mathbb{Z}$, called the time axis, $\mathbb{W}$ a set called the signal space, and $\mathfrak{B}$ a subset of $\mathbb{W}^\mathbb{T}$ called the behavior. Here $\mathbb{W}^\mathbb{T}$ is the notation for the collection of all maps from $\mathbb{T}$ to $\mathbb{W}$.
}
\end{def1} 

This definition of dynamical systems in behavioral theory presents a general framework for common system representations like ordinary differential equations, state space models and transfer functions because they all define functions that describe the time dependence of a trajectory evolution in a signal space. We call any collection of time depending functions the \textit{behavior} of the given models. Generally speaking, this framework makes a formal distinction between a system (its behavior) and its representations. 

In the rest of this paper, we will only consider systems evolving over discrete time: $\mathbb{T}:=\mathbb{N}_0$ and initialized at $t=0$. 

A simple discrete-time example is given to illustrate the definition above.
\begin{exmp}
Consider a linear discrete-time state space system given as
\begin{equation} \label{nonsss}
\Sigma:\left\{
\begin{aligned}
x(t+1)&=Ax(t)+Bu(t);\\
y(t)&=Cx(t), \hspace{4mm} x(0)\in X_0,
\end{aligned}\right.
\end{equation} 
with $x(t)\in X\subseteq \mathbb{R}^n, u(t)\in U\subseteq \mathbb{R}^p,y(t)\in Y\subseteq \mathbb{R}^k$ and $X_0\subseteq X$. Then, the \emph{full} behavior or the \emph{input/state/output} behavior of (\ref{nonsss}) is given as
$$\mathfrak{B}_{\mbox{i/s/o}}:=\{(u,x,y)\in (U\times X\times Y)^\mathbb{T}\mid (\ref{nonsss}) \mbox{ is satisfied}\}.$$
The variable $x$ is considered as a latent variable, therefore the \emph{manifest} behavior or the \emph{input/output} behavior is given by
\begin{equation}\nonumber
\begin{aligned}
\mathfrak{B}_{\mbox{i/o}}:=\{(u,y)\in (U\times Y)^\mathbb{T}\mid \exists &x\in X^\mathbb{T}\\ &\mbox{s.t. }(u,x,y)\in \mathfrak{B}_{\mbox{i/s/o}}\}.
\end{aligned}
\end{equation}

When looking at the classical systems and control field, specifications are usually defined over the input/output behavior. And within the domain of formal methods, we often only consider the specifications over the \emph{output} behavior. In our work, we tackle the second ``simple'' view on specifications over the output behavior and develop theory for it. Hence, the output behavior $y\in Y^\mathbb{T}$ that we are interested in is defined as 
\begin{center}
$\mathfrak{B}^\mathbf{y}:=\Pi_\mathbf{y}\left(\mathfrak{B}_{\mbox{i/o}}\right)$
\end{center}
with $\Pi_\mathbf{y}$ a projection map taking $(u,y)\in (U\times Y)^\mathbb{T}$ to $y\in Y^\mathbb{T}$.\QED
\end{exmp}  

Behavioral theory treats system interconnections as variable sharing. This is different from
classical control theory, which views interconnection as channels through which outputs of one system are imposed as inputs to another system. 
\begin{def1}
\textit{Let $\Sigma_1=(\mathbb{T},\mathbb{W}_1\times \mathbb{C},\mathfrak{B}_1)$ and $\Sigma_2=(\mathbb{T},\mathbb{W}_2\times \mathbb{C},\mathfrak{B}_2)$ be two dynamical systems. Then the interconnection of $\Sigma_1$ and $\Sigma_2$, denoted by $\Sigma=\Sigma_1\times \Sigma_2$, is the system $\Sigma=(\mathbb{T},\mathbb{W}_1\times \mathbb{W}_2\times \mathbb{C},\mathfrak{B})$ with $\mathfrak{B}=\{(w_1,w_2,c):\mathbb{T}\rightarrow \mathbb{W}_1\times \mathbb{W}_2\times \mathbb{C}\mid (w_1,c)\in \mathfrak{B}_1,(w_2,c)\in \mathfrak{B}_2\}.$ 
}
\end{def1}

\begin{figure}[!h]
\centering
\includegraphics[width=0.8\columnwidth]{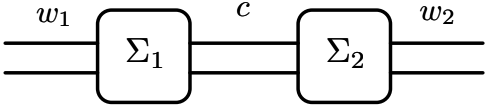}
\caption{Partial interconnection structure.}
\label{partialstructure}
\end{figure}

This kind of interconnection structure is called partial interconnection~\cite{mutsaers2012control} as shown in Fig \ref{partialstructure}. We can see that $c\in \mathbb{C}^\mathbb{T}$ is shared by both $\Sigma_1$ and $\Sigma_2$ while $w_1\in \mathbb{W}_1^\mathbb{T}$ only belongs to $\Sigma_1$ and $w_2\in \mathbb{W}_2^\mathbb{T}$ only belongs to $\Sigma_2$. Especially, if both $\mathbb{W}_1$ and $\mathbb{W}_2$ are empty, the full interconnection structure is obtained and $\mathfrak{B}=\mathfrak{B}_1\cap \mathfrak{B}_2$.

\begin{figure}[!h]
\centering
\includegraphics[width=0.5\columnwidth]{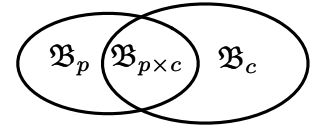}
\caption{Controlled behavior $\mathfrak{B}_{p\times c}$ is the intersection of the plant and controller behavior $\mathfrak{B}_p$ and $\mathfrak{B}_c$.}
\label{cb}
\end{figure}

In the behavioral theory, control is best understood through interconnections and variable sharing, rather than signal or information transmitting in classical system theory. From the behavior point of view, control means restricting the behavior of a system, namely, the plant, through the interconnection with another system, namely, the controller~\cite{willems2013introduction,willems2007behavioral}. As shown in Fig \ref{cb},  the control problem aims to find a controller $\Sigma_c=(\mathbb{T},\mathbb{W},\mathfrak{B}_c)$, with the behavior $\mathfrak{B}_c$, that after the interconnection with the plant $\Sigma_p=(\mathbb{T},\mathbb{W},\mathfrak{B}_p)$, with the behavior $\mathfrak{B}_p$, results in the controlled system $\Sigma_{p\times c}:=\Sigma_p\times \Sigma_c=(\mathbb{T},\mathbb{W},\mathfrak{B}_p\cap \mathfrak{B}_c)$~\cite{mutsaers2012control}. Here, we define a well-posed controller $\Sigma_c$ for $\Sigma_p$.  
\begin{def1}\label{well-posed}
\textit{Given a plant $\Sigma_p=(\mathbb{T},\mathbb{W},\mathfrak{B}_p)$, we say that a system $\Sigma_c=(\mathbb{T},\mathbb{W},\mathfrak{B}_c)$ is a well-posed controller for $\Sigma_p$ if the following conditions are satisfied:\\
1. $\mathfrak{B}_{p\times c}:=\mathfrak{B}_p\cap \mathfrak{B}_c\neq \{\emptyset\};$\\
2. For any initial state, there exists unique continuation in $\mathfrak{B}_{p\times c}$.
}
\end{def1}

A well-posed controller $\Sigma_c$ for $\Sigma_p$ is denoted as $\Sigma_c\in\mathfrak{C}(\Sigma_p)$ and all well-posed controllers make up the well-posed controller set $\mathfrak{C}(\Sigma_p)$.

\subsection{DAE $\&$ control refinement} 
Consider a linear DAE system $\Sigma:=(E,A,B,C)$ defined as
\begin{equation}\label{cdaeee}
\Sigma:\left\{
\begin{aligned}
Ex(t+1)&=Ax(t)+Bu(t);\\
y(t)&=Cx(t),\hspace{4mm} x(0)\in X_0,
\end{aligned}\right.
\end{equation}
with $x(t)\in X\subseteq \mathbb{R}^n, u(t)\in U\subseteq \mathbb{R}^p,y(t)\in Y\subseteq \mathbb{R}^k$ as its state, input and output, respectively. $E,A\in \mathbb{R}^{n\times n},B\in \mathbb{R}^{n\times p}$ and $C\in \mathbb{R}^{k\times n}$ are constant matrices. We assume, without loss of generality, that rank$(B)=p$ and rank$(C)=k$. For the special case that $E$ is nonsingular, the DAE system (\ref{cdaeee}) can be transformed into a \emph{standard} state space system and we also call it a standard DAE system. 

We refer to $\Sigma$ as the \emph{concrete} DAE system if it is the DAE for which we would like to develop the controller. That is the DAE that actually represents the physical system in which we are interested.  

The manifest behavior of (\ref{cdaeee}) is given as 
\begin{equation}\label{behavior1}
\begin{aligned}
\mathfrak{B}_{\Sigma}:=\{(u,y)\in (U\times Y)^\mathbb{T}&\mid \exists x\in X^\mathbb{T}\\\mbox{ s.t. }&(u,x,y)\mbox{ satisfies } (\ref{cdaeee})\}.
\end{aligned}
\end{equation}

An \emph{abstract} linear DAE system $\Sigma_a:= (E_a,A_a,B_a,C_a)$ is defined as
\begin{equation} \label{adaeeee}
\Sigma_a:\left\{
\begin{aligned}
E_ax_a(t+1)&=A_ax_a(t)+B_au_a(t);\\
y_a(t)&=C_ax_a(t),\hspace{4mm} x_a(0)\in X_{a0},
\end{aligned}\right.
\end{equation} 
with $x_a(t)\in X_a\subseteq \mathbb{R}^m,u_a(t)\in U_a\subseteq \mathbb{R}^q,y_a(t)\in Y_a\subseteq \mathbb{R}^k$. In this paper, we consider an abstract DAE system $\Sigma_a$ that is of the same dimension or simpler than the concrete DAE system $\Sigma$, i.e., $m\leq n$. Similarly, the input/output behavior of the abstract DAE system $\Sigma_a$ is derived as
\begin{equation}\nonumber
\begin{aligned}
\mathfrak{B}_{\Sigma_a}:=\{(u_a,y_a)\in (U_a\times Y_a)^\mathbb{T}&\mid \exists x_a\in X_a^\mathbb{T}\\\mbox{ s.t. }&(u_a,x_a,y_a)\mbox{ satisfies } (\ref{adaeeee})\}.
\end{aligned}
\end{equation}

Of interest to us is how can we refine a well-posed controller $\Sigma_{c_a}$ for $\Sigma_a$ to attain a well-posed controller $\Sigma_c$ for $\Sigma$ such that the output behavior of the two controlled systems is exactly the same or the distance between them is bounded within the error $\varepsilon$, which we will formulate in the sequel. First we introduce the notions of exact and approximate control refinement. 
\begin{def1}\label{exact_refine_def}
\textit{(Exact control refinement). Let $\Sigma_a$ and $\Sigma$ be the abstract and concrete systems, respectively. We say that controller $\Sigma_{c}$ refines the controller $\Sigma_{c_a}\in \mathfrak{C}(\Sigma_a)$ if $\mathfrak{B}_{\Sigma_a\times \Sigma_{c_a}}\neq \emptyset$ and $\mathfrak{B}^\mathbf{y}_{\Sigma\times \Sigma_{c}}\subseteq \mathfrak{B}^\mathbf{y}_{\Sigma_a\times \Sigma_{c_a}}$.}
\end{def1}

The exact control refinement requires that the controlled output behavior of the abstract and the concrete systems is exactly the same, while the approximate control refinement only requires that the distance between their controlled output behavior is bounded within the error $\varepsilon$. Recall the notation of $\mathcal{E}_\varepsilon$~\cite{moor2001robust}, the approximate control refinement is defined by requiring the output behavior of the controlled concrete system to lie in the $\varepsilon$-expansion of the output behavior of the controlled abstract system. Hence, as a contrast, the approximate control refinement is defined as follows: 
\begin{def1}\label{appro_refine_def}
\textit{(Approximate control refinement). Let $\Sigma_a$ and $\Sigma$ be the abstract and concrete systems, respectively. We say that controller $\Sigma_{c}$ refines the controller $\Sigma_{c_a} \in \mathfrak{C}(\Sigma_a)$ if $\mathfrak{B}_{\Sigma_a\times \Sigma_{c_a}}\neq \emptyset$ and $\mathfrak{B}^\mathbf{y}_{\Sigma\times \Sigma_{c}}\subseteq \mathcal{E}_\varepsilon\big(\mathfrak{B}^\mathbf{y}_{\Sigma_a\times \Sigma_{c_a}}\big)$.} 
\end{def1}
\subsection{Problem statement}
As proposed in the introduction, for a given concrete DAE that actually represents the physical system, it is usually difficult to develop a controller for it directly due to the combination of algebraic equations and high state dimensions. Hence, according to Definition \ref{exact_refine_def} and Definition \ref{appro_refine_def}, we pose the problem that how can we tackle this by developing control refinement approaches. That is, given any well-posed controller of the abstract DAE system, for which controller design is much easier than that of the concrete DAE, can we always refine that to attain a well-posed controller for the concrete model and how can we develop the refined controller. 

First of all, before tackling the control refinement problems, we need to consider the problem that what is a well-posed controller $\Sigma_{c_a}$ for the abstract DAE system $\Sigma_a$. Whereafter, for any such $\Sigma_{c_a}$, further we question whether it is possible to refine $\Sigma_{c_a}$ to $\Sigma_{c}$ via Definition \ref{exact_refine_def}. That is, we consider the problem whether for every well-posed controller $\Sigma_{c_a}$ designed for $\Sigma_a$, there always exists a well-posed controller $\Sigma_{c}$ for $\Sigma$ such that the two controlled systems have the same output behavior. The exact control refinement problem can be formulated as follows.

\begin{problem}\label{problem1}
\textit{(Exact control refinement). Let $\Sigma_a$ and $\Sigma$ be the abstract and concrete systems, respectively. For any well-posed controller $\Sigma_{c_a}\in \mathfrak{C}(\Sigma_a)$, refine $\Sigma_{c_a}$ to $\Sigma_{c}$, s.t. $\Sigma_{c}\in \mathfrak{C}(\Sigma)$ and $\mathfrak{B}^\mathbf{y}_{\Sigma\times \Sigma_{c}}\subseteq \mathfrak{B}^\mathbf{y}_{\Sigma_a\times \Sigma_{c_a}}$.}
\end{problem}
\vspace{1mm}

Unlike exact control refinement, approximate relationships, which do allow for the possibility of error, will certainly provide more freedom for controller design. Therefore, as a contrast, further we consider the approximate control refinement problem between the concrete model $\Sigma$ and its approximation $\Sigma_a$. Under the same settings for exact cases, we question how to refine a well-posed controller $\Sigma_{c_a}$ to attain a well-posed controller $\Sigma_{c}$ such that the distance between the output behavior of the two controlled systems is bounded by a tolerated error $\varepsilon$. Recall the notation of $\mathcal{E}_\varepsilon$~\cite{moor2001robust} and Definition \ref{appro_refine_def}, the approximate control refinement problem can be formulated.
\begin{problem}\label{problem2}
\textit{(Approximate control refinement). Let $\Sigma_a$ and $\Sigma$ be the abstract and concrete systems, respectively. For any well-posed controller $\Sigma_{c_a}\in \mathfrak{C}(\Sigma_a)$, refine $\Sigma_{c_a}$ to $\Sigma_{c}$, s.t. $\Sigma_{c}\in \mathfrak{C}(\Sigma)$ and $\mathfrak{B}^\mathbf{y}_{\Sigma\times \Sigma_{c}}\subseteq \mathcal{E}_\varepsilon\big(\mathfrak{B}^\mathbf{y}_{\Sigma_a\times \Sigma_{c_a}}\big)$.}
\end{problem}

\section{Models, Behavior $\&$ Properties}
\noindent Since we deal with DAE models, in the very beginning of this section, we introduce the basic concepts and properties about linear DAE systems to get some insights of these so-called DAEs. Whereafter, we present the definition of transition systems that enables us to treat these systems in a more general framework. Subsequently, we introduce the notions of (bi)simulation relations and approximate simulation relations, which will be used later to develop approaches for exact and approximate control refinement, respectively. The (bi)simulation relations propose new notions of system equivalence while approximate simulation relations introduce system relationships that bound the distance between the output behavior of two systems. Finally, simulation functions that are widely used for hierarchical control of standard state space systems are proposed. 

\subsection{Linear DAE systems}
In this subsection, we recall the DAE system $\Sigma=(E,A,B,C)$ defined by (\ref{cdaeee}) with the input/output behavior given as (\ref{behavior1}).

A special case that is of interest in this work, and for which the associated Weierstrass Canonical form is developed, is the so-called \textit{regular} DAE systems with regular matrix pencils $(E,A)$ defined as  
\begin{def1}
\textit{Let $E,A\in \mathbb{R}^{m\times n}$. The matrix pencil $(E,A)$ is called \emph{regular} if $m=n$ and the characteristic polynomial p defined by
\begin{equation}
p(\lambda)=det(\lambda E-A)
\end{equation} is not the zero polynomial. A matrix pencil that is not regular is called singular.}
\end{def1}

In our work, we assume that all the DAE systems are regular and the theorem of Weierstrass canonical form is introduced.
\begin{thm}\label{wcf}
\textit{
(Weierstrass canonical form)~\cite{kunkel2006differential}. Let the matrix pencil $(E,A)$ of (\ref{cdaeee}) be regular, then there exists non-singular matrices $P$ and $Q$ that transform the system to Weierstass canonical form, 
\begin{equation}
\tilde{E}:=PEQ=\left[\begin{matrix}I_{n_1}&0\\0&N
\end{matrix}\right],\tilde{A}:=PAQ=\left[\begin{matrix}J&0\\0&I_{n_2}
\end{matrix}\right], \nonumber
\end{equation}
\begin{equation}
\tilde{B}:=PB=\left[\begin{matrix}
B_1\\ B_2
\end{matrix}\right],
\tilde{C}:=CQ=\left[\begin{matrix}
C_1&C_2
\end{matrix}\right],
\end{equation}
\begin{equation}
Q^{-1}x=\left[\begin{matrix}
x_1\\ x_2
\end{matrix}\right],
Q^{-1}x_0=\left[\begin{matrix}
x_{10}\\ x_{20}
\end{matrix}\right] \nonumber
\end{equation}
\noindent where $x_1\in \mathbb{R}^{n_1}, x_2\in \mathbb{R}^{n_2}$ and $n_1+n_2=n$. $J\in \mathbb{R}^{n_1\times n_1}$ is a matrix in Jordan canonical form and $N\in \mathbb{R}^{n_2\times n_2}$ is a nilpotent matrix also in Jordan canonical form and the nilpotency $\mu$ of $N$ is called the  index of the system, denoted by $\mu=\ind(\Sigma)$.}
\end{thm}

Then we use the Weierstrass canonical form to give some insights of the DAE systems. DAE systems always show some freedom in the choice of the next states $x(t+1)$, which is the nondeterminism of the DAE systems. We propose this Weierstrass canonical form in this paper because it introduces a way of working with DAE systems, especially it is useful to derive the state evolutions and the related output trajectories. According to Theorem \ref{wcf}, the DAE system (\ref{cdaeee}) is decomposed into two subsystems. One is a standard state-space subsystem and another one is an anti-causal subsystem, denoted by $\Sigma^c$ and $\Sigma^a$, respectively.\\
\indent The causal subsystem has the following representation
\begin{equation}
\Sigma^c:\left\{
\begin{matrix}
\begin{aligned}
x_{1}(t+1) &=Jx_{1}(t)+B_1u(t);\\
y_{1}(t) &=C_{1}x_{1}(t),\hspace{0.5cm} x_{1}(0)\in X_{10}.
\end{aligned} 
\end{matrix} \label{causal}
\right.
\end{equation}

\noindent The anti-causal subsystem is as follows
\begin{equation}
\Sigma^a:\left\{
\begin{matrix}
\begin{aligned}
Nx_{2}(t+1) &=x_{2}(t)+B_{2}u(t);\\
y_{2}(t) &=C_{2}x_{2}(t),\hspace{0.5cm} x_{2}(0)\in X_{20}.
\end{aligned} 
\end{matrix} \label{acausal}
\right.
\end{equation}
\noindent Therefore, the output behavior of the system (\ref{cdaeee}) is
\begin{equation}
\begin{aligned}
\mathfrak{B}_{\Sigma}^\mathbf{y}:= \Pi_\mathbf{y}(\mathfrak{B}_{\Sigma})=\{y_1+y_2\mid y_1 &\mbox{ output of } (\ref{causal}),\\ &y_2\mbox{ output of } (\ref{acausal})\}.
\end{aligned}
\end{equation}

\indent After decomposing system (\ref{cdaeee}) into two subsystems, the time domain properties of the system are considered. At time $t$, the state responses for subsystems (\ref{causal}) and (\ref{acausal}) are
\begin{equation}
\begin{matrix}
\begin{aligned}
x_{1}(t)&=J^tx_1(0)+\sum_{\tau=0}^{t-1}J^{t-\tau-1}B_{1}u(\tau); \\
 x_{2}(t)&=-\sum_{\tau=0}^{\mu-1}N^{\tau}B_{2}u(t+\tau).
\end{aligned} 
\end{matrix} \label{state}
\end{equation}
\indent With the initial condition $x_1(0)=0$, the associated output of the system is defined as
\begin{equation}
\begin{matrix}
\begin{aligned}
y_{1}(t)&=\sum_{\tau=0}^{t-1}C_1J^{t-\tau-1}B_{1}u(\tau); \\
 y_{2}(t)&=-\sum_{\tau=0}^{\mu-1}C_2N^{\tau}B_{2}u(t+\tau);\\
 y(t)&=y_1(t)+y_2(t), \hspace{0.5cm} t\in \mathbb{N}_0.
\end{aligned} 
\end{matrix} \label{output}
\end{equation}

From system output (\ref{output}), it can be clearly seen that the DAE system contains an anti-causal part since $y_2$ depends on the future input and the anti-causality horizon is determined by the system index $\mu$.

After giving the input-output relationship, the reachability of the DAE systems is investigated that will make sense in the system transformation later. Other properties like observability and stability can be found in~\cite{Xingangthesis, dai1989singular, debeljkovic2011stability}.
 
The DAE system (\ref{cdaeee}) is reachable if and only if both of the subsystems $\Sigma^c$ and $\Sigma^a$ are reachable~\cite{stykel2004gramian}. The reachability of the causal subsystem (\ref{causal}) is the same as reachability for standard state space systems~\cite{hespanha2009linear}. The reachability of the entire DAE system (\ref{cdaeee}) is defined as follows.
\begin{def1}
\textit{The DAE system (\ref{cdaeee}) is said to be reachable if for any $x_{f}\in \mathbb{R}^{n}$, there exists $t_1\in \mathbb{T}$ and an input function $u(t)$ that steers the zero initial state $x(0)=0$ to $x(t_1)=x_f$ in some finite time $t_1$.
}
\end{def1}

This definition means that under the reachability assumption, a control input that drives the zero initial state to desired position in finite time can always be found.

Consider the state responses of anti-causal subsystem given in equation (\ref{state}), the following equation is derived.
\begin{equation}\nonumber
x_2(t)=-
\underbrace{\left[\begin{matrix} B_2&NB_2& \cdots & N^{\mu-1}B_2 \end{matrix}\right]}_{\mathcal{R}_\mu}
\left[\begin{matrix}
u(t)\\u(t+1)\\\vdots \\u(t+\mu-1)
\end{matrix}
\right]
\end{equation}
with $\mathcal{R}_\mu\in \mathbb{R}^{n_2\times p\mu}$. Consider the anti-causal subsystem (\ref{acausal}), in order to find an input that steers the zero initial state $x_2(0)=0$ to $x_2(t)$, $\mathcal{R}_\mu$ should have full row rank, i.e., rank ($\mathcal{R}_\mu)=n_2$ and this also means that $n_2\leq p\mu$. Therefore, the following proposition is developed and refer to~\cite{stykel2004gramian} for the proof.
\begin{prop} \label{conmatrix}
\textit{The DAE system $\Sigma$ is reachable if and only if the causal subsystem $\Sigma^c$ and the anti-causal subsystem $\Sigma^a$ are both reachable, or equivalently the reachability matrices
\begin{equation}
\mathcal{R}_{c}=\left[\begin{matrix} B_1&JB_1& \cdots & J^{n_1-1}B_1 \end{matrix}\right]\in \mathbb{R}^{n_1\times pn_1}, \nonumber
\end{equation}
\begin{equation}
\mathcal{R}_{\mu}=\left[\begin{matrix} B_2&NB_2& \cdots & N^{\mu-1}B_2 \end{matrix}\right]\in \mathbb{R}^{n_2\times p\mu} \nonumber
\end{equation}
both have full row rank, i.e., 
\begin{equation}
rank (\mathcal{R}_c)=n_1, rank (\mathcal{R}_\mu)=n_2.\nonumber
\end{equation}
}
\end{prop}

Dually, the observability for the DAE system (\ref{cdaeee}) can be developed similarly and is omitted here. In this paper, we only deal with control refinement for DAE systems that are both reachable and observable. 

\subsection{Transition systems}
\begin{def1} \label{transition}
\textit{A transition system $\Sigma=(X,U,X_0,\rightarrow,Y,\mathcal{O})$ consists of:
\begin{itemize}
\item a set of states X,
\item a set of inputs U,
\item a set of initial states $X_0\subseteq X$,
\item a transition relation $\rightarrow \subseteq X\times U\times X$,
\item a set of outputs Y,
\item an output map $\mathcal{O}:X\rightarrow Y$.
\end{itemize}
}
\end{def1}

Given any initial state $x(0)\in X_0$, we construct the infinite sequence of transitions
\begin{equation}\nonumber
x(0)\xrightarrow{u(0)}x(1)\xrightarrow{u(1)}x(2)\xrightarrow{u(2)}x(3)\xrightarrow{u(3)}\cdots
\end{equation}
such that $(x(t),u(t),x(t+1))\in \rightarrow$ over discrete time $\mathbb{T}:=\mathbb{N}_0$. This infinite
sequence of transitions defines the state trajectory. The related output trajactory is
\begin{equation}\nonumber
y(0)\rightarrow y(1)\rightarrow y(2)\rightarrow y(3)\cdots.
\end{equation}
All these trajectories $(u,y)$ make up the manifest behavior of the transition system $\Sigma$ and the behavior is initialized at $t=0$. 

A system is called \textit{blocking} if there is a state $x\in X$ from which no further transitions are possible, i.e., $x$ has no $u$-successor for any $u\in U$. A system is called \textit{non-blocking} if the set of successors of every $x\in X$ is nonempty, i.e., $\forall x, \exists (u,x^\prime)$ such that $(x,u,x^\prime)\in \rightarrow$ and $x^\prime$ is called an $u$-successor of $x$.

A system is called \textit{deterministic} if for any state $x\in  X$ and any input $u\in U$, $x\xrightarrow{u} x^\prime$ and $x\xrightarrow{u} x^{\prime\prime}$ implies $x^\prime=x^{\prime\prime}$. Therefore, a system is called deterministic if given any state $x\in  X$ and any input $u\in U$, there exists at most one $u$-successor (there may be none)~\cite{tabuada}. A system is called \textit{nondeterministic} if it is not deterministic.

In order to quantify the desired precision we need a metric on the set of outputs, so the definition of metric transition system is introduced.
\begin{def1}
\textit{~\cite{metric} A transition system $\Sigma=(X,U,X_0,\rightarrow,Y,\mathcal{O})$ is called a metric transition system if $(Y,d_Y)$ is a metric space, where $d_Y:Y\times Y\rightarrow \mathbb{R}_0^+$.
}
\end{def1}

DAE systems can be treated in this transition system framework and the following example is considered.
\begin{exmp}\label{transitionexam}
The DAE system (\ref{cdaeee}) is also a transition system $\Sigma=(X,U,X_{0},\rightarrow,Y,\mathcal{O})$ with:
\textit{
\begin{itemize}
\item the set of states is $X\subseteq \mathbb{R}^{n}$,
\item the set of inputs is $U\subseteq \mathbb{R}^p$,
\item the set of initial values is $X_{0}\subseteq X$,
\item the transition relation $\rightarrow:=(x,u,x^+)\in X\times U\times X\mbox{ s.t. } Ex^+=Ax+Bu$,
\item the set of outputs is $Y\subseteq \mathbb{R}^k$,
\item the output map is $\mathcal{O}:y=Cx$.\QED
\end{itemize}
}
\end{exmp}

\begin{rem} 
This is a nondeterministic system because some of the next states $x(t+1)$ are free to choose due to the singularity of $E$ and can be resolved by designing controllers to remove the nondeterminism.  
\end{rem}
 
\subsection{Simulation relations}
Essentially, a simulation relation of $\Sigma_1$ by $\Sigma_2$ is a relation on the states of the systems that describes how to select transitions of $\Sigma_2$ in order to match the transitions of $\Sigma_1$ and to produce the same output behavior as $\Sigma_1$. 

\begin{def1}
\textit{Consider two systems $\Sigma_1$ and $\Sigma_2$, a relation $\mathcal{R}\subseteq X_1\times X_2$ is called a simulation relation of $\Sigma_1$ by $\Sigma_2$, if the following conditions are satisfied:\\
1. $\forall (x_1,x_2)\in \mathcal{R}$, we have $\mathcal{O}_1(x_1)=\mathcal{O}_2(x_2)$,\\
2. $\forall (x_1,x_2)\in \mathcal{R}$ and transitions $x_1\xrightarrow[1]{u_1} x_1^\prime$, there exists a transtion $x_2\xrightarrow[2]{u_2} x_2^\prime$, such that $(x_1^\prime,x_2^\prime)\in \mathcal{R}$.\\
We say that $\Sigma_1$  is simulated by $\Sigma_2$, denoted by $\Sigma_1\preceq \Sigma_2$, if there exists a simulation relation $\mathcal{R}$ of $\Sigma_1$ by $\Sigma_2$ and in addition $\forall x_{10}\in X_{10}, \exists x_{20}\in X_{20}$ such that $(x_{10},x_{20})\in \mathcal{R}$.
}
\end{def1}
\vspace{1mm}

If a relation $\mathcal{R}\subseteq X_1\times X_2$ is a simulation relation of $\Sigma_1$ by $\Sigma_2$ and in addition its inverse $\mathcal{R}^{-1}\subseteq X_2\times X_1$ is a simulation relation of $\Sigma_2$ by $\Sigma_1$, we call $\mathcal{R}$ a bisimulation relation between $\Sigma_1$ and $\Sigma_2$.  
\begin{def1}
\textit{Consider two systems $\Sigma_1$ and $\Sigma_2$, a relation $\mathcal{R}\subseteq X_1\times X_2$ is called a bisimulation relation between $\Sigma_1$ and $\Sigma_2$, if the following conditions are satisfied:\\
1. $\forall (x_1,x_2)\in \mathcal{R}$, we have $\mathcal{O}_1(x_1)=\mathcal{O}_2(x_2)$,\\
2. $\forall (x_1,x_2)\in \mathcal{R}$ and transitions $x_1\xrightarrow[1]{u_1} x_1^\prime$, there exists a transtion $x_2\xrightarrow[2]{u_2} x_2^\prime$, such that $(x_1^\prime,x_2^\prime)\in \mathcal{R}$.\\
3. $\forall (x_1,x_2)\in \mathcal{R}$ and transitions $x_2\xrightarrow[2]{u_2} x_2^\prime$, there exists a transtion $x_1\xrightarrow[1]{u_1} x_1^\prime$, such that $(x_1^\prime,x_2^\prime)\in \mathcal{R}$.\\
We say that $\Sigma_1$ and $\Sigma_2$ are bisimilar, denoted by $\Sigma_1\cong \Sigma_2$, if there exists a bisimulation relation $\mathcal{R}$ between $\Sigma_1$ and $\Sigma_2$ and in addition $\forall x_{10}\in X_{10}, \exists x_{20}\in X_{20}$ s.t. $(x_{10},x_{20})\in \mathcal{R}$ and $\forall x_{20}\in X_{20}, \exists x_{10}\in X_{10}$ s.t. $(x_{10},x_{20})\in \mathcal{R}$.
}
\end{def1}
\vspace{1mm}

The notion of approximate simulation relation is obtained by relaxing the equality of the output behavior. Instead of the identical behavior, approximate simulation relation requires that the distance between the output behavior remains bounded. The definition of an approximate simulation relation is given as follows. 
\begin{def1}
\textit{Consider two systems $\Sigma_1$ and $\Sigma_2$, let $\varepsilon\geq0$, a relation $\mathcal{R}_\varepsilon\subseteq X_1\times X_2$ is called an approximate simulation relation of $\Sigma_1$ by $\Sigma_2$ of precision $\varepsilon$, if the following conditions are satisfied:\\
1. $\forall (x_1,x_2)\in \mathcal{R}_\varepsilon$, we have $d(\mathcal{O}_1(x_1),\mathcal{O}_2(x_2))\leq\varepsilon$,\\
2. $\forall (x_1,x_2)\in \mathcal{R}_\varepsilon$ and transitions $x_1\xrightarrow[1]{u_1} x_1^\prime$, there exists a transition $x_2\xrightarrow[2]{u_2} x_2^\prime$, such that $(x_1^\prime,x_2^\prime)\in \mathcal{R}_\varepsilon$.\\
We say that $\Sigma_2$ approximately simulates $\Sigma_1$ with precision $\varepsilon$, denoted by $\Sigma_1\preceq_{\varepsilon} \Sigma_2$, if there exists an approximate simulation relation $\mathcal{R}_\varepsilon$ of $\Sigma_1$ by $\Sigma_2$ and in addition $\forall x_{10}\in X_{10}, \exists x_{20}\in X_{20}$ s.t. $(x_{10},x_{20})\in \mathcal{R}_\varepsilon$.
}
\end{def1}
\vspace{1mm}

After giving the notions of (bi)simulation relations and approximate simulation relations, the property of transitivity~\cite{tabuada} of these notions is considered because it can be used to construct (approximate) simulation relations for DAE systems later. 
\begin{prop}\label{transitivity}
\textit{(Transitivity). Let $\mathcal{R}_{\varepsilon_1}$ be an approximate simulation relation from $\Sigma_1$ to $\Sigma_2$ and $\mathcal{R}_{\varepsilon_2}$ be an approximate simulation relation from $\Sigma_2$ to $\Sigma_3$. In addition, $\forall x_{10}\in X_{10}, \exists x_{20}\in X_{20}$ s.t. $(x_{10},x_{20})\in \mathcal{R}_{\varepsilon_1}$ and $\forall x_{20}\in X_{20}, \exists x_{30}\in X_{30}$ s.t. $(x_{20},x_{30})\in \mathcal{R}_{\varepsilon_2}$. Then, we can conclude that  
\begin{equation}\nonumber
\begin{aligned}
\mathcal{R}_{\varepsilon_1+\varepsilon_2}:=\{(x_1,x_3)\mid &\exists x_2\\&\mbox{ s.t. } (x_1,x_2)\in \mathcal{R}_{\varepsilon_1}, (x_2,x_3)\in \mathcal{R}_{\varepsilon_2}\}.
\end{aligned}
\end{equation}
is an approximate simulation relation from $\Sigma_1$ to $\Sigma_3$, and in addition $\forall x_{10}\in X_{10}, \exists x_{30}\in X_{30}$ s.t. $(x_{10},x_{30})\in \mathcal{R}_{\varepsilon_1+\varepsilon_2}$.
}
\end{prop}
\vspace{1mm}
 
(Approximate) simulation relations between two deterministic transitions systems imply a class of functions called \textit{interfaces}, which are proposed in hierarchical control for standard state space systems~\cite{main}. An interface maps actions of the first system and the current states of the two systems to the actions for the second system such that the states of the two systems belong to the related (approximate) simulation relation under the parallel state evolutions.
\begin{def1}\label{interfacedef}
\textit{(Interface). Let $\Sigma_1=(X_1,U_1,X_{10},\rightarrow_1,Y,\mathcal{O}_1)$ and $\Sigma_2=(X_2,U_2,X_{20},\rightarrow_2,Y,\mathcal{O}_2)$ be two deterministic transition systems with an approximate simulation relation $\mathcal{R}_\varepsilon$ from $\Sigma_1$ to $\Sigma_2$. Then $\mathcal{F}:U_1\times X_1\times X_2\mapsto U_2$ is an interface related to $\mathcal{R}_\varepsilon$, if the following conditions are satisfied:\\
1. for every $(x_1,x_2)\in \mathcal{R}_\varepsilon$, we have that $x_1{\xrightarrow{u_1}}_1x_1^\prime$ in $\Sigma_1$ implies $x_2{\xrightarrow{u_2}}_2x_2^\prime$ in $\Sigma_2$ with $u_2=\mathcal{F}(u_1,x_1,x_2)$, satisfying $(x_1^\prime,x_2^\prime)\in \mathcal{R}_\varepsilon$.\\ 
2. $\forall x_{20}\in X_{20}, \exists x_{10}\in X_{10}$ s.t. $(x_{10},x_{20})\in \mathcal{R}_\varepsilon$. 
}
\end{def1}
\vspace{1mm} 

In fact, simulation relations always imply the output behavior inclusion. We can conclude the following proposition and refer to~\cite{tabuada} for the similar proof.
\begin{prop}\label{behaviorrelations}
\textit{Let $\Sigma_1=(X_1,U_1,X_{10},\rightarrow_1,Y,\mathcal{O}_1)$ and $\Sigma_2=(X_2,U_2,X_{20},\rightarrow_2,Y,\mathcal{O}_2)$ be two metric transition systems. Then, the following implications hold:
\begin{equation}\nonumber
\begin{aligned}
&\Sigma_1\preceq \Sigma_2 \Longrightarrow \mathfrak{B}_{\Sigma_1}^\mathbf{y}\subseteq \mathfrak{B}_{\Sigma_2}^\mathbf{y},\\
&\Sigma_1\cong \Sigma_2 \Longrightarrow \mathfrak{B}_{\Sigma_1}^\mathbf{y}=\mathfrak{B}_{\Sigma_2}^\mathbf{y},\\
&\Sigma_1\preceq_{\varepsilon} \Sigma_2 \Longrightarrow \mathfrak{B}_{\Sigma_1}^\mathbf{y}\subseteq \mathcal{E}_\varepsilon\big(\mathfrak{B}_{\Sigma_2}^\mathbf{y}\big).
\end{aligned}
\end{equation}
}  
\end{prop}
\vspace{1mm}
 
We now consider the following example to get more insights of the simulation relations between DAE systems and their behavior.

\begin{exmp}
Consider a concrete DAE system $\Sigma=(E,A,B,C)$ with 
\begin{equation}\nonumber
E=\begin{bsmallmatrix}
1&0&0\\0&0&1\\0&0&0
\end{bsmallmatrix},A=\begin{bsmallmatrix}
-1&0&0\\0&1&0\\0&0&1
\end{bsmallmatrix},B=\begin{bsmallmatrix}
1\\2\\0
\end{bsmallmatrix},C=\begin{bsmallmatrix}
0.2\\0.5\\1
\end{bsmallmatrix}^T.
\end{equation}
$\Sigma$ is not a minimal realization because it is observable but not reachable by checking the observability and reachability matrices. It is also a transition system denoted by $\Sigma=(X,U,X_{0},\rightarrow,Y,\mathcal{O})$, where $X_0\subseteq X$ and $X,U,Y$ are respectively subsets of $\mathbb{R}^3,\mathbb{R},\mathbb{R}$. The transition relation is $\rightarrow:=(x,u,x^+)\in X\times U\times X\mbox{ s.t. } Ex^+=Ax+Bu$, and the output map is $\mathcal{O}:y=Cx$. Based on Silverman-Ho algorithm~\cite{dai1989singular}, we choose an abstract DAE system $\Sigma_a=(E_a,A_a,B_a,C_a)$ that is the minimal realization of $\Sigma$ and 
\begin{equation}\nonumber
E_a=\begin{bsmallmatrix}
1&0\\0&0
\end{bsmallmatrix},A_a=\begin{bsmallmatrix}
-1&0\\0&1
\end{bsmallmatrix},B_a=\begin{bsmallmatrix}
1\\1
\end{bsmallmatrix},C_a=\begin{bsmallmatrix}
0.2\\1
\end{bsmallmatrix}^T.
\end{equation}
Similarly, $\Sigma_a$ is also a transition system denoted by $\Sigma_a=(X_a,U_a,X_{a0},\rightarrow_a,Y,\mathcal{O}_a)$, where $X_{a0}\subseteq X_a$ and $X_a,U_a,Y$ are respectively subsets of $\mathbb{R}^2,\mathbb{R},\mathbb{R}$. The transition relation is $\rightarrow_a:=(x_a,u_a,x_a^+)\in X_a\times U_a\times X_a\mbox{ s.t. } E_ax_a^+=A_ax_a+B_au_a$, and the output map is $\mathcal{O}_a:y_a=C_ax_a$. 

Subsequently, 
\begin{equation}\nonumber
\mathcal{R}:=\{(x_a,x)\mid x=\mathcal{H}x_a, x_a\in X_a, x\in X\}
\end{equation}
is a bisimulation relation between $\Sigma_a$ and $\Sigma$, where
\begin{equation}\nonumber
\mathcal{H}=\begin{bsmallmatrix}
1&0\\0&2\\0&0
\end{bsmallmatrix}.
\end{equation}
Then, we consider the two requirements of bisimulation relations. For any $(x_a,x)\in\mathcal{R}$, we have $Cx=C\mathcal{H}x_a=C_ax_a$. For any $(x_a,x)\in\mathcal{R}$ with $x_a=(x_{a1},x_{a2})^T$, we have $x\in X$ denoted by $x=(x_1,x_2,0)^T$ with $x_1=x_{a1}, x_2=2x_{a2}$. Then, consider the transition $x_a{\xrightarrow{u_a}}_a x_a^\prime$ in $\Sigma_a$ with $u_a=-x_{a2},x_a^\prime=(-x_{a1}-x_{a2},x_{a2}^\prime)$ and $x_{a2}^\prime$ is free to choose. Take the action $u=u_a$ in $\Sigma$, then the transition $x\xrightarrow{u} x^\prime$ results in the next state $x^\prime=(-x_1-0.5x_2,x_2^\prime)$ with $x_2^\prime=2x_{a2}^\prime$. Therefore, $(x_a^\prime,x^\prime)\in\mathcal{R}$. Conversely, for any transition $x\xrightarrow{u} x^\prime$ in $\Sigma$ with $u=-0.5x_2, x^\prime=(-x_1-0.5x_2,x_2^\prime)$ and $x_2^\prime$ is free to choose. Take the action $u_a=u$ in $\Sigma_a$, then the transition $x_a{\xrightarrow{u_a}}_a x_a^\prime$ results in the next state $x_a^\prime=(-x_{a1}-x_{a2},x_{a2}^\prime)$ with $x_{a2}^\prime=0.5x_2^\prime$. We also have $(x_a^\prime,x^\prime)\in\mathcal{R}$ and finally we have proven that this $\mathcal{R}$ is a bisimulation relation.

In addition, for any $x_0\in X_0$ denoted by $(x_{10},x_{20},0)^T$, there exists $x_{a0}=(x_{a10},x_{a20})^T=(x_{10},0.5x_{20})^T\in X_{a0}$ s.t. $(x_{a0},x_0)\in \mathcal{R}$. Conversely, for any $x_{a0}\in X_{a0}$, there exists $x_0=\mathcal{H}x_{a0}\in X_{a0}$ s.t. $(x_{a0},x_0)\in \mathcal{R}$. Therefore, we can conclude that $\Sigma_a\cong \Sigma$. Consequently, according to Proposition \ref{behaviorrelations}, we obtain $\mathfrak{B}_{\Sigma_a}^\mathbf{y}=\mathfrak{B}_{\Sigma}^\mathbf{y}$ with the behavior initialized at $t=0$. \QED
\end{exmp}

\subsection{Simulation functions}
In this subsection, we focus on the definitions of simulation functions, which will define the corresponding approximate simulation relations directly. In fact, a simulation function is a positive function that bounds the distance between the output behavior and non-increasing under the parallel evolution of the systems.
\begin{def1}\label{bridge}~\cite{bridge}
\textit{A function $\mathcal{S}:X_1\times X_2\rightarrow \mathbb{R}^{+}\cup \{+\infty\}$ is called a simulation function of $\Sigma_1$ by $\Sigma_2$ if its sub-level sets are closed, and for all $(x_1,x_2)\in X_1\times X_2$:
\begin{eqnarray} \label{generalsim}
\mathcal{S}(x_1,x_2)\geq \max \Bigg\{d(\mathcal{O}_1(x_1),\mathcal{O}_2(x_2)),\nonumber \\
\sup_{\substack{x_1^\prime \in T_1(x_1,u_1)}} \inf_{\substack{x_2^\prime \in T_2(x_2,u_2)}} \mathcal{S}(x_1^\prime,x_2^\prime)\Bigg\}. 
\end{eqnarray}
}
\end{def1}
\vspace{3mm}

\begin{prop}
\textit{let $\mathcal{S}$ be a simulation function of $\Sigma_1$ by $\Sigma_2$, then, for all $\varepsilon\geq 0$,
\begin{eqnarray}
\mathcal{R}_\varepsilon =\{(x_1,x_2)\in X_1\times X_2\mid \mathcal{S}(x_1,x_2)\leq \varepsilon\} \nonumber
\end{eqnarray}
is an approximate simulation relation of $\Sigma_1$ by $\Sigma_2$ of precision $\varepsilon$.
}
\end{prop}

Particularly, the zero set (if exists) of a simulation function is a simulation relation.

\section{Exact control refinement for DAEs}
\noindent In this section, we focus on exact control refinement for DAE systems via the notions of (bi)simulation relations. We first consider the exact control refinement for standard DAE systems and after that, we introduce a kind of systems called driving variable (DV) systems, which are in standard state space forms. The DV systems that are bisimilar or behaviorally equivalent to the related DAE systems, provide ease in control refinement for DAE systems. Subsequently, we develop algorithms to transform DAE systems into DV systems and vice versa. We show that the DAE systems and the related DV systems are bisimilar and behaviorally equivalent. All these procedures will benefit the exact control refinement for DAEs, which will be presented in the end of this section. 

\subsection{Control refinement for standard DAE systems}
Consider the concrete and abstract DAE systems $\Sigma:=(I_n,A,B,C)$ and $\Sigma_a:=(I_m,A_a,B_a,C_a)$ in standard state space forms by setting $E=I_n$ and $E_a=I_m$ in (\ref{cdaeee}) and (\ref{adaeeee}). Both $\Sigma_a$ and $\Sigma$ are deterministic. The control refinement between $\Sigma_a$ and $\Sigma$ is developed based on a simulation relation $\mathcal{R}$ from $\Sigma_a$ to $\Sigma$, and in addition $\forall x_{0}\in X_{0}, \exists x_{a0}\in X_{a0}$ s.t. $(x_{a0},x_{0})\in \mathcal{R}$. The simulation relation $\mathcal{R}$ and the initialization conditions imply that there exists an interface from $\Sigma_a$ to $\Sigma$ as shown in Definition \ref{interfacedef}. As a result, we have the following lemma. 
\begin{lem}\label{standarddaeexact}
\textit{Let $\Sigma_a$ and $\Sigma$ be the standard abstract and concrete DAE systems written in two metric transition systems $\Sigma_a=(X_a,U_a,X_{a0},\rightarrow_a,Y,\mathcal{O}_a)$ and $\Sigma=(X,U,X_{0},\rightarrow,Y,\mathcal{O})$. $\mathcal{R}$ is a simulation relation from $\Sigma_a$ to $\Sigma$ and $\mathcal{F}:U_a\times X_a\times X\mapsto U$ is a related interface. Then, for any controller $\Sigma_{c_a}\in \mathfrak{C}(\Sigma_a)$, the controller $\Sigma_c:(\Sigma_a\times \Sigma_{c_a})\times \mathcal{F}$ refines $\Sigma_{c_a}$ such that all initial states $x_0\in X_0$ have continuation in $\mathfrak{B}_{\Sigma\times \Sigma_{c}}$ and $\mathfrak{B}^\mathbf{y}_{\Sigma\times \Sigma_{c}}\subseteq \mathfrak{B}^\mathbf{y}_{\Sigma_a\times \Sigma_{c_a}}$.}
\end{lem}

The proof of Lemma \ref{standarddaeexact} can be developed based on the properties of simulation relations and the related interfaces.      
\subsection{DAE to DV conversion}
Usually, it is difficult to deal with DAE systems directly. In this subsection, we introduce a new kind of system representation called driving variable (DV) system~\cite{weiland1991theory} that is in state space form. We will investigate that whether the DAE system and the related DV system are bisimilar or behaviourally equivalent.

First of all, consider the following system with the same state space as (\ref{cdaeee}), and a new free driving input $s(t)$. The outputs of the system $(u(t),y(t))\in U\times Y$ are the vectorized input and output of DAE system (\ref{cdaeee}). This kind of system is called a driving variable system~\cite{weiland1991theory} and denoted by $\Sigma_{\mbox{\scriptsize{DV}}}:=(A_d,B_d,C_d,D_d,C_u,D_u)$,
\begin{equation} \label{dv_2}
\Sigma_{\mbox{\scriptsize{DV}}}:\left\{
\begin{aligned}
x(t+1)&=A_dx(t)+B_ds(t);\\
\left(\begin{matrix}
u(t)\\y(t)
\end{matrix}
 \right)&=C_dx(t)+D_ds(t),\hspace{0.16cm}x(0)\in X_0,
\end{aligned}\right.
\end{equation} 
where $u(t)=C_u x(t)+D_u s(t)$ and $x(t)\in X\subseteq \mathbb{R}^n, s(t)\in S\subseteq  \mathbb{R}^p, u(t)\in U \subseteq\mathbb{R}^p, y(t)\in Y\subseteq \mathbb{R}^k, X_0\subseteq X$. Hence, the behavior of the DV system (\ref{dv_2}) is defined as
\begin{equation}\nonumber
\begin{aligned}
\mathfrak{B}_{\Sigma_{\mbox{\scriptsize{DV}}}}:=\{(u,y)\in \left(U\times Y\right)^\mathbb{T}\mid \exists (x,s)&\in \left(X\times S\right)^\mathbb{T},\\& \mbox{ s.t.}(\ref{dv_2}) \mbox{ holds}\}.
\end{aligned}
\end{equation}

If $\mathfrak{B}_{\Sigma_{\mbox{\scriptsize{DV}}}}=\mathfrak{B}_{\Sigma}$, we say that the DAE system (\ref{cdaeee}) and the DV system (\ref{dv_2}) are behaviorally equivalent. This notion is used to establish the connection between DAE systems and DV systems, that is to rewrite DAE systems as DV systems and back. 

%
%

Any concrete DAE system (\ref{cdaeee}) that is reachable can be rewritten as the related concrete DV system. The conversion formulation is developed based on the kernel and right inverse of $\left[\begin{matrix}
E&-B
\end{matrix}\right]$ and is shown as Algorithm \ref{daetodvalgorithm}. Refer to Appendix \uppercase\expandafter{\romannumeral2} for the computation details.

\begin{algorithm}
\caption{DAE $\rightarrow$ DV conversion algorithm}\label{daetodvalgorithm}
\begin{algorithmic}[1]
\INPUT{A DAE system $\Sigma:=(E,A,B,C)$ with $\begin{bmatrix}E&B\end{bmatrix}$ full row rank.}
\OUTPUT{A DV system $\Sigma_{\mbox{\scriptsize{DV}}}:=(A_d,B_d,C_d,D_d,C_u,D_u)$ such that $u(t)=C_ux(t)+D_us(t)$.}
\Procedure{DAE $\rightarrow$ DV transformation}{}
\State Let $M:=\begin{bmatrix}E&-B\end{bmatrix}$;
\State Compute the right inverse $M^+$ of $M$;
\State Let $B_{\mathcal{N}}$ be a matrix s.t. $\ker M=\im B_{\mathcal{N}}$;
\State Decompose $M^+=\begin{bmatrix}
M_x\\M_u
\end{bmatrix}
$, $B_{\mathcal{N}}=\begin{bmatrix}
B_{\mathcal{N}}^x\\B_{\mathcal{N}}^u\end{bmatrix}
$;
\State Set $A_d=M_xA$, $B_d=B_{\mathcal{N}}^x$;
\State Set $C_u=M_uA$, $D_u=B_{\mathcal{N}}^u$;
\State Stack matrices to obtain $C_d=\begin{bmatrix}
C_u\\C
\end{bmatrix}
$,$D_d=\begin{bmatrix}
D_u\\0
\end{bmatrix}
$.
\BState \textbf{end}
\EndProcedure
\end{algorithmic}
\end{algorithm}

Hence, the concrete DAE system (\ref{cdaeee}) can be rewritten into the following concrete DV system $\Sigma_{\mbox{\scriptsize{DV}}}=(A_d,B_d,C_d,D_d,C_u,D_u)$ based on Algorithm 1 and we present the expressions of $u$ and $y$ separately.
\begin{equation} \label{dv2new}
\Sigma_{\mbox{\scriptsize{DV}}}:\left\{
\begin{aligned}
x(t+1)&=A_dx(t)+B_ds(t);\\
u(t)&=C_ux(t)+D_us(t);\\
y(t)&=Cx(t),\hspace{5mm}x(0)\in X_0,
\end{aligned}\right.
\end{equation} 
with $x(t)\in X\subseteq \mathbb{R}^n, s(t)\in S\subseteq  \mathbb{R}^p, u(t)\in U \subseteq\mathbb{R}^p, y(t)\in Y\subseteq \mathbb{R}^k, X_0\subseteq X$. In our work, since we are interested in the output behavior of the DAE system. Hence, for the related DV system, solely $y(t)$ is regarded as the output and $u(t)$ is treated as an intermediate that represents the input given to the corresponding DAE system. Therefore, consider the output map $\mathcal{O}:y=Cx$ solely, the DV system is also a transition system 
$\Sigma_{\mbox{\scriptsize{DV}}}=(X,S,X_0,\rightarrow_{\mbox{\tiny{DV}}},Y,\mathcal{O})$.
\subsection{DV to DAE conversion}
In the previous subsection, we rewrote DAE system (\ref{cdaeee}) into a DV system (\ref{dv2new}). Conversely, in this subsection, we develop an algorithm to rewrite the DV system (\ref{dv2new}) back into the DAE system $\Sigma$. In addition, $s(t)$ can be expressed by $x(t), x(t+1)$ and $u(t)$. The algorithm is developed based on the singular value decomposition (SVD) of ${\begin{bmatrix}B_d^T&D_u^T\end{bmatrix}}^T$ and is shown as Algorithm \ref{dvtodaealgorithm}. For the computation details, we refer to Appendix \uppercase\expandafter{\romannumeral2}.

\begin{algorithm}
\caption{DV $\rightarrow$ DAE conversion algorithm}\label{dvtodaealgorithm}
\begin{algorithmic}[1]
\INPUT{A DV system $\Sigma_{\mbox{\scriptsize{DV}}}:=(A_d,B_d,C_d,D_d,C_u,D_u)$.}
\OUTPUT{A DAE system $\Sigma:=(E,A,B,C)$ and a matrix $W$ such that $s(t)=W\begin{bmatrix}x(t+1)^T&u(t)^T&x(t)^T\end{bmatrix}^T$.}
\Procedure{DV $\rightarrow$ DAE transformation}{}
\State Decompose $C_d:=\begin{bmatrix}C_u\\C\end{bmatrix}$, $D_d:=\begin{bmatrix}D_u\\0\end{bmatrix}$;
\State Let $\begin{bmatrix}A_d\\C_u\end{bmatrix}=\mathcal{Q}$, $\begin{bmatrix}B_d\\D_u\end{bmatrix}=\mathcal{P}$;
\State Develop the SVD $\mathcal{P}=U\Sigma V^T$ with $\Sigma=\begin{bmatrix}\bar{\Sigma}^T&0\end{bmatrix}^T$;
\State Decompose $U$ into the fist $p$ and last $n$ columns: $U=\begin{bmatrix}U_p&U_n\end{bmatrix}$;
\State Decompose $U_n^T$ into the fist $n$ and last $p$ columns: $U_n^T=\begin{bmatrix}U_1^T&U_2^T\end{bmatrix}$;
\State Set $E=U_1^T, A=U_n^T\mathcal{Q}, B=-U_2^T$;
\State Construct $W=V\bar{\Sigma}^{-1}\begin{bmatrix}U_p^T&-U_p^T\mathcal{Q}\end{bmatrix}$.
\BState \textbf{end}
\EndProcedure
\end{algorithmic}
\end{algorithm}

We know that $\Sigma$ and $\Sigma_{\mbox{\scriptsize{DV}}}$ are behaviorally equivalent via behavior approach. In fact, as shown in Section \uppercase\expandafter{\romannumeral3}, bisimilarity always implies output behavioral equivalence. Hence, the following proposition that proposes a the stronger relationship of bisimilarity between a DAE system and its related DV system is concluded. The proof is given in Appendix \uppercase\expandafter{\romannumeral1}.
\begin{thm}\label{pairwise}
\textit{$\Sigma$ and $\Sigma_{\mbox{\scriptsize{DV}}}$ are bisimilar, and consequently 
$\mathfrak{B}_{\Sigma}^\mathbf{y}=\mathfrak{B}_{\Sigma_{\mbox{\scriptsize{DV}}}}^\mathbf{y}.
$
}
\end{thm} 

\subsection{Main result: exact control refinement for DAEs}
\begin{figure}[!h]
\centering
\includegraphics[width=0.8\columnwidth]{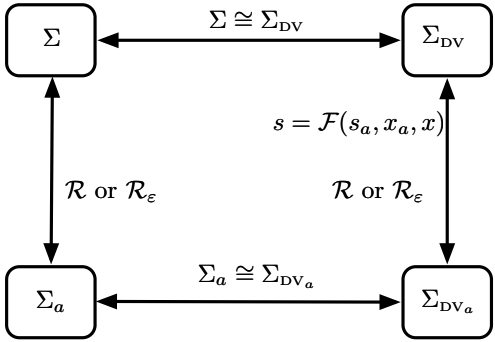}
\caption{Connection between DAE and DV framework.}
\label{newstructure}
\end{figure}

In this subsection, we focus on the solution of Problem 1. We show that if there exists a simulation relation $\mathcal{R}$ from $\Sigma_a$ to $\Sigma$, in addition $\forall x_{0}\in X_{0}, \exists x_{a0}\in X_{a0}$ s.t. $(x_{a0},x_{0})\in \mathcal{R}$. Then for any well-posed controller $\Sigma_{c_a}\in \mathfrak{C}(\Sigma_a)$, we can always refine $\Sigma_{c_a}$ to attain a controller $\Sigma_{c}$ for $\Sigma$ such that $\Sigma_{c}\in \mathfrak{C}(\Sigma)$ and $\mathfrak{B}^\mathbf{y}_{\Sigma_a\times \Sigma_{c_a}}\subseteq \mathfrak{B}^\mathbf{y}_{\Sigma\times \Sigma_c}$. This claim can be proved directly by developing an exact control refinement approach that will be presented in the remaining of this section. This approach is developed based on our previous results of conversions between DAE systems and the related DV systems. The general framework is shown as Figure \ref{newstructure} and it illustrates the connections between the DAE framework and the DV framework.  

As shown in Figure \ref{newstructure}, in the horizontal direction, the DAE framework and the DV framework are connected by bisimulation relations and in the vertical direction, the abstract models and the concrete models are connected by the simulation relations or approximate simulation relations and the related interfaces. $\Sigma_{\mbox{\scriptsize{DV}}_a}=(A_{da},B_{da},C_{da},D_{da},C_{u_a},D_{u_a})$ is the abstract DV system defined as (\ref{dv1new}) 
\begin{equation} \label{dv1new}
\Sigma_{\mbox{\scriptsize{DV}}_a}:\left\{
\begin{aligned}
x_a(t+1)&=A_{da}x(t)+B_{da}s_a(t);\\
u_a(t)&=C_{u_a}x_a(t)+D_{u_a}s_a(t);\\
y_a(t)&=C_ax_a(t),\hspace{5mm}x_a(0)\in X_{a0},
\end{aligned}\right.
\end{equation}  
where $u_a(t)=C_{u_a}x_a(t)+D_{u_a}s_a(t)$ and $x_a(t)\in X_a\subseteq \mathbb{R}^m, s_a(t)\in S_a\subseteq  \mathbb{R}^q, u_a(t)\in U_a \subseteq\mathbb{R}^q, y_a(t)\in Y\subseteq \mathbb{R}^k, X_{a0}\subseteq X_a$. The abstract DV system is also a transition system $\Sigma_{\mbox{\scriptsize{DV}}_a}=(X_a,S_a,X_{a0},\rightarrow_{\mbox{\tiny{DV}}_a},Y,\mathcal{O}_a)$. The behavior of the abstract DV system (\ref{dv1new}) is defined as
\begin{equation}\nonumber
\begin{aligned}
\mathfrak{B}_{\Sigma_{\mbox{\scriptsize{DV}}_a}}:=\{(u_a,y_a)&\in \left(U_a\times Y\right)^\mathbb{T}\mid \exists (x_a,s_a)\\ &\in \left(X_a\times S_a\right)^\mathbb{T}, \mbox{ s.t.}(\ref{dv1new}) \mbox{ holds}\}.
\end{aligned}
\end{equation}

The four systems $\Sigma, \Sigma_{\mbox{\scriptsize{DV}}}, \Sigma_{\mbox{\scriptsize{DV}}_a}$ and $\Sigma_a$ build up our framework in Figure \ref{newstructure} for developing control refinement approaches. According to the transitivity of (approximate) simulation relations and the initialization conditions given by Proposition \ref{transitivity}, we can conclude that there exists a simulation relation $\mathcal{R}^d$ from $\Sigma_{\mbox{\scriptsize{DV}}_a}$ to $\Sigma_{\mbox{\scriptsize{DV}}}$, in addition $\forall x_{0}\in X_{0}, \exists x_{a0}\in X_{a0}$ s.t. $(x_{a0},x_{0})\in \mathcal{R}^d$. This also indicates that there exists an interface from $\Sigma_{\mbox{\scriptsize{DV}}_a}$ to $\Sigma_{\mbox{\scriptsize{DV}}}$. 

Before giving the exact control refinement approach, we first consider what is a well-posed controller $\Sigma_{c_a}$ for the abstract DAE system $\Sigma_a$. Let us consider the following controller $\Sigma_{c_a}$ defined by a linear DAE. Although we define a linear controller here, this can also be extended to nonlinear controllers. 
\begin{equation}\label{DAEcontroller}
\Sigma_{c_a}:E_cx_a(t+1)=A_cx_a(t)+B_cu_a(t),
\end{equation}
with $E_c,A_c\in \mathbb{R}^{n_c\times m}$ and $B_c\in \mathbb{R}^{n_c\times q}$. The interconnected system $\Sigma_a\times \Sigma_{c_a}$ is derived as
\begin{equation}\label{DAEcontrollerinter}
\begin{bmatrix}E_a\\E_c\end{bmatrix}x_a(t+1)=\begin{bmatrix}A_a\\A_c\end{bmatrix}x_a(t)+\begin{bmatrix}B_a\\B_c\end{bmatrix}u_a(t).
\end{equation}
(\ref{DAEcontrollerinter}) can be rewritten as 
\begin{equation}\label{DAErewrite11}
\begin{bmatrix}E_a&-B_a\\E_c&-B_c\end{bmatrix}\begin{bmatrix}x_a(t+1)\\u_a(t)\end{bmatrix}=\begin{bmatrix}A_a\\A_c\end{bmatrix}x_a(t).
\end{equation}

The controller $\Sigma_{c_a}$ is admissible if (\ref{DAErewrite11}) is nonblocking, that is, for any $x_a(t)\in X_a$, there always exists a pair $(x_a(t+1),u_a(t))$ such that (\ref{DAErewrite11}) holds. In addition, if the pair $(x_a(t+1),u_a(t))$ is unique for such $x_a(t)$, which means the controlled output behavior is unique once initialized, we say that $\Sigma_{c_a}\in \mathfrak{C}(\Sigma_a)$. Subsequently, we develop the following lemma by referring to~\cite{abadir2005matrix}, which discusses the solutions of the matrix equality $Ax=b$.
\begin{lem}\label{well-posed}
\textit{The controller $\Sigma_{c_a}$ is admissible with infinite solutions if and only if 
\begin{equation}\label{lemma9}
rank\begin{pmatrix}
\begin{bsmallmatrix}
E_a&B_a\\E_c&B_c
\end{bsmallmatrix}
\end{pmatrix}
=rank\begin{pmatrix}
\begin{bsmallmatrix}
E_a&B_a&A_a\\E_c&B_c&A_c
\end{bsmallmatrix}
\end{pmatrix}< m+q.
\end{equation}
The controller $\Sigma_{c_a}\in \mathfrak{C}(\Sigma_a)$ if and only if 
\begin{equation}\label{lemma99}
rank\begin{pmatrix}
\begin{bsmallmatrix}
E_a&B_a\\E_c&B_c
\end{bsmallmatrix}
\end{pmatrix}
=rank\begin{pmatrix}
\begin{bsmallmatrix}
E_a&B_a&A_a\\E_c&B_c&A_c
\end{bsmallmatrix}
\end{pmatrix}= m+q.
\end{equation}
}
\end{lem}
\vspace{2mm}

In order to start with a well-posed controller $\Sigma_{c_a}$, we know that the augmented matrix on the left of (\ref{DAErewrite11}) should have full column rank, then it has a left inverse. Thus, multiplying this left inverse by the left on both sides of (\ref{DAErewrite11}), the controlled system is an autonomous system in standard state space form (\ref{closedloop1}).
\begin{equation}\label{closedloop1}
\Sigma_a\times \Sigma_{c_a}:\left\{
\begin{aligned}
x_a(t+1)&=\mathcal{A}_ax_a(t);\\
y_a(t)&=C_ax_a(t),\hspace{4mm} x_a(0)\in X_{a0},
\end{aligned}\right.
\end{equation}
with $u_a(t)=\mathcal{B}_ax_a(t)$. 
 
After giving the conditions for well-posed controllers, we first consider the exact control refinement from $\Sigma_{\mbox{\scriptsize{DV}}}$ to $\Sigma$ and develop the following theorem.
\begin{thm}\label{bisimulationtheorem}
\textit{Let $\Sigma$ be the concrete DAE system as (\ref{cdaeee}), $\Sigma_{\mbox{\scriptsize{DV}}}$ is the related DV system as (\ref{dv2new}) such that $\Sigma_{\mbox{\scriptsize{DV}}}\cong \Sigma$. Then, for any control strategy $s(t)$ of $\Sigma_{\mbox{\scriptsize{DV}}}$, the controller  
\begin{equation}\label{refinelaw}
\Sigma_c:\left\{
\begin{aligned}
B_{d}^Tx(t+1)&=B_{d}^TA_{d}x(t)+B_{d}^TB_{d}s(t);\\
u(t)&=C_ux(t)+D_us(t)
\end{aligned}\right.
\end{equation}
refines $s(t)$ such that $\Sigma_{\mbox{\scriptsize{DV}}}$ and $\Sigma$ have the same controlled output behavior.}
\end{thm} 

The proof of Theorem \ref{bisimulationtheorem} is given in Appendix \uppercase\expandafter{\romannumeral1}. This theorem also proposes an approach to stabilize a DAE system. 

Subsequently, we switch the problem around and consider the exact control refinement from $\Sigma_a$ to $\Sigma_{\mbox{\scriptsize{DV}}_a}$. For a well-posed controller $\Sigma_{c_a}$ of $\Sigma_a$ given as (\ref{DAEcontroller}) with a closed loop $\Sigma_a\times \Sigma_{c_a}$ defined as (\ref{closedloop1}), using the expression of $s_a(t)=W_1x_a(t+1)+W_2u_a(t)+W_3x_a(t)$ as shown in Algorithm 2, we derive
\begin{equation}\nonumber
s_a(t)=(W_1\mathcal{A}_a+W_2\mathcal{B}_a+W_3)x_a(t).
\end{equation}

Then, we can conclude the following theorem about the control refinement from $\Sigma_a$ to $\Sigma_{\mbox{\scriptsize{DV}}_a}$.  
\begin{thm}\label{bisimulationtheorem1}
\textit{Let $\Sigma_a$ be the abstract DAE system as (\ref{adaeeee}), $\Sigma_{\mbox{\scriptsize{DV}}_a}$ is the related DV system as (\ref{dv1new}) such that $\Sigma_{\mbox{\scriptsize{DV}}_a}\cong \Sigma_a$. Then, for any well-posed controller $\Sigma_{c_a}$ of $\Sigma_a$ defined as (\ref{DAEcontroller}) together with the closed loop $\Sigma_a\times \Sigma_{c_a}$ defined as (\ref{closedloop1}), the control strategy  
\begin{equation}\label{refinelaw1}
s_a(t):=\mathcal{T}(x_a(t))=(W_1\mathcal{A}_a+W_2\mathcal{B}_a+W_3)x_a(t)
\end{equation}
refines $\Sigma_{c_a}$ such that $\Sigma_{\mbox{\scriptsize{DV}}_a}$ and $\Sigma_a$ have the same controlled output behavior.}
\end{thm} 

The proof of Theorem \ref{bisimulationtheorem1} is given in Appendix \uppercase\expandafter{\romannumeral1}. According to theorems \ref{bisimulationtheorem} and \ref{bisimulationtheorem1}, we can develop the approach for exact control refinement from the abstract DAE system $\Sigma_a$ to the concrete DAE system $\Sigma$. First of all, the simulation relation $\mathcal{R}^d$ from $\Sigma_{\mbox{\scriptsize{DV}}_a}$ to $ \Sigma_{\mbox{\scriptsize{DV}}}$ and the initialization conditions imply that there exists an interface $s=\mathcal{F}(s_a,x_a,x)$ between them. Finally, we derive the following theorem as a solution for Problem 1.
\begin{thm}\label{solution1}
\textit{Let $\Sigma$ and $\Sigma_a$ be the given concrete and abstract DAE systems defined as (\ref{cdaeee}) and (\ref{adaeeee}), respectively. $\mathcal{R}$ is a simulation relation from $\Sigma_a$ to $\Sigma$, and in addition $\forall x_0\in X_0, \exists x_{a0}\in X_{a0}\mbox{ s.t. } (x_{a0},x_0)\in \mathcal{R}$. Then for any $\Sigma_{c_a}\in\mathfrak{C}(\Sigma_a)$ defined as (\ref{DAEcontroller}), the controller  
\begin{equation}\label{solution1law}
\Sigma_c:\left\{
\begin{aligned}
B_d^Tx(t+1)&=B_d^TA_dx(t)+B_d^TB_ds(t);\\
u(t)&=C_ux(t)+D_us(t),
\end{aligned}\right.
\end{equation}
with $s(t)=\mathcal{F}(\mathcal{T}(x_a(t)),x_a(t),x(t))$, refines $\Sigma_{c_a}$ such that $\Sigma_c\in \mathfrak{C}(\Sigma_c)$ and $\mathfrak{B}^\mathbf{y}_{\Sigma\times \Sigma_{c}}\subseteq \mathfrak{B}^\mathbf{y}_{\Sigma_a\times \Sigma_{c_a}}$.}
\end{thm}

The proof of Theorem \ref{solution1} is generally based on the proofs of Theorem \ref{bisimulationtheorem}, Theorem \ref{bisimulationtheorem1} and Lemma \ref{standarddaeexact}. 

In the end of this section, we consider the following simple example as an interpretation. 
\begin{exmp}\label{example2}
Consider the given concrete DAE system $\Sigma=(E,A,B,C)$ with 
\begin{equation}\nonumber
E=\begin{bsmallmatrix}
1&0&0\\0&0&1\\0&0&0
\end{bsmallmatrix},A=\begin{bsmallmatrix}
-1&0&0\\0&1&0\\0&0&1
\end{bsmallmatrix},B=\begin{bsmallmatrix}
1\\1\\1
\end{bsmallmatrix},C=\begin{bsmallmatrix}
0.1\\0.2\\0.5
\end{bsmallmatrix}^T.
\end{equation}

with $x(t)=\begin{bmatrix}
x_1(t)&x_2(t)&x_3(t)
\end{bmatrix}^T$. Employ Algorithm 1 regarding the DAE to DV conversion, $\Sigma_{\mbox{\scriptsize{DV}}}=(A_d,B_d,C_d,D_d,C_u,D_u)$ is derived as
\begin{equation}\label{exmp2dv}
\Sigma_{\mbox{\scriptsize{DV}}}:\left\{\begin{aligned}
x(t+1)&=\begin{bsmallmatrix}
-1&0&-1\\0&0&0\\0&1&-1
\end{bsmallmatrix}x(t)+\begin{bsmallmatrix}
0\\-1\\0
\end{bsmallmatrix}s(t)\\
u(t)&=\begin{bsmallmatrix}
0&0&-1\end{bsmallmatrix}x(t);\\
y(t)&=\begin{bsmallmatrix}
0.1&0.2&0.5
\end{bsmallmatrix}x(t),\hspace{5mm} x(0)\in X_0.
\end{aligned}\right.
\end{equation}
As we can see, in this example, the second state $x_2(t+1)$ is free to choose in $\Sigma$, which is also revealed in $\Sigma_{\mbox{\scriptsize{DV}}}$. Looking at $\Sigma_{\mbox{\scriptsize{DV}}}$, $x_2(t+1)$ is just determined by the current driving input $s(t)$. Once the control strategy for $s(t)$ is determined, the free state $x_2(t+1)$ will be restricted. Afterwards, we can refine the control strategy of $s(t)$ to a control strategy for $\Sigma$ and the nondeterminism of $x_2(t+1)$ can be removed. Consider the following abstract DV system $\Sigma_{\mbox{\scriptsize{DV}}_a}$, which is similar to $\Sigma_{\mbox{\scriptsize{DV}}}$. 
\begin{equation}\label{exmp2dva}
\Sigma_{\mbox{\scriptsize{DV}}_a}:\left\{\begin{aligned}
x_a(t+1)&=\begin{bsmallmatrix}
-1&0&-1\\0&0&0\\0&1&-1
\end{bsmallmatrix}x_a(t)+\begin{bsmallmatrix}
0\\-1\\0
\end{bsmallmatrix}s_a(t)\\
u_a(t)&=\begin{bsmallmatrix}
0&0&-1\end{bsmallmatrix}x_a(t);\\
y_a(t)&=\begin{bsmallmatrix}
0.1&0.2&0.5
\end{bsmallmatrix}x_a(t),\hspace{4mm} x(a0)\in X_{a0}.
\end{aligned}\right.
\end{equation}
$\mathcal{R}^d=\{(x_a,x)\in X_a\times X\mid x_a-x=0\}$ is a simulation relation from $\Sigma_{\mbox{\scriptsize{DV}}_a}$ to $\Sigma_{\mbox{\scriptsize{DV}}}$. In addition $\forall x_0\in X_0, \exists x_{a0}\in X_{a0}\mbox{ s.t. } (x_{a0},x_0)\in \mathcal{R}_d$. Then $s=\mathcal{F}(s_a,x_a,x)=s_a+K(x-x_a)$ is the related interface, where $K$ is a stabilizing gain for $\Sigma_{\mbox{\scriptsize{DV}}}$. 
 
Afterwards, according to algorithm 2, the abstract DAE system $\Sigma_a=(E_a,A_a,B_a,C_a)$ is developed with
\begin{equation}\nonumber
E_a=\begin{bsmallmatrix}
1&0&0\\0&0&1\\0&0&0
\end{bsmallmatrix},A_a=\begin{bsmallmatrix}
-1&0&-1\\0&1&-1\\0&0&-1
\end{bsmallmatrix},B_a=\begin{bsmallmatrix}
0\\0\\-1
\end{bsmallmatrix},C_a=\begin{bsmallmatrix}
0.1\\0.2\\0.5
\end{bsmallmatrix}^T.
\end{equation}
$\mathcal{R}=\{(x_a,x)\in X_a\times X\mid x_a-x=0\}$ is a simulation relation from $\Sigma_a$ to $\Sigma$ based on the transitivity of relations in Proposition \ref{transitivity}, in addition $\forall x_0\in X_0, \exists x_{a0}\in X_{a0}\mbox{ s.t. } (x_{a0},x_0)\in \mathcal{R}$. 
Subsequently, consider a well-posed controller $\Sigma_{c_a}\in \mathfrak{C}(\Sigma_a)$ defined as
\begin{equation}\nonumber
\Sigma_{c_a}:\begin{bsmallmatrix}
0&-1&0
\end{bsmallmatrix}x_a(t+1)=\begin{bsmallmatrix}
0.5&-1.4&3
\end{bsmallmatrix}x_a(t)+u_a(t),
\end{equation}
and the closed loop is
\begin{equation}\nonumber
\Sigma_a\times \Sigma_{c_a}:\left\{
\begin{aligned}
x_a(t+1)&=\begin{bsmallmatrix}
-1&0&-1\\-1.5&2.4&-4\\0&1&-1
\end{bsmallmatrix}x_a(t);\\
y_a(t)&=\begin{bsmallmatrix}
0.1&0.2&0.5
\end{bsmallmatrix}x_a(t),\hspace{4mm} x(a0)\in X_{a0},
\end{aligned}\right.
\end{equation}
 with $u_a(t)=\begin{bsmallmatrix}
0&0&-1
\end{bsmallmatrix}x_a(t)$ and $\Sigma_a\times \Sigma_{c_a}$ is stable. Then based on Theorem \ref{bisimulationtheorem1}, we derive that
\begin{equation}\nonumber
s_a(t)=\begin{bsmallmatrix}
1.5&-2.4&\hspace{1mm}4
\end{bsmallmatrix}x_a(t).
\end{equation}
We obtain a controlled abstract DV system that is the same as $\Sigma_a\times \Sigma_{c_a}$ by applying $s_a(t)$ to (\ref{exmp2dva}). Whereafter, the refined controller $\Sigma_c$ for $\Sigma$ is derived based on Theorem \ref{solution1} together with the interface $s=\mathcal{F}(s_a,x_a,x)$.
\begin{equation}\nonumber
\Sigma_c: \left\{
\begin{aligned}
\begin{bsmallmatrix}
0&-1&0
\end{bsmallmatrix}x(t+1)&=Kx(t)+(\begin{bsmallmatrix}
1.5&-2.4&\hspace{1mm}4
\end{bsmallmatrix}-K)x_a(t);\\
u(t)&=\begin{bsmallmatrix}
0&0&-1
\end{bsmallmatrix}x(t).
\end{aligned}\right.
\end{equation}

Consider the special simulation relation $\mathcal{R}$ in this example, in addition $\forall x_0\in X_0, \exists x_{a0}\in X_{a0}\mbox{ s.t. }(x_{a0},x_0)\in \mathcal{R}$. We start from this initial pair $(x_{a0},x_0)\in \mathcal{R}$ that results in $(x_a(t),x(t))\in \mathcal{R},t\in \mathbb{N}_0$. Thus $x_a(t)=x(t),t\in \mathbb{N}_0$. Eventually, based on $x_a(t)=x(t)$, the closed loop $\Sigma\times \Sigma_c$ is derived as
\begin{equation}\nonumber
\Sigma\times \Sigma_{c}:\left\{
\begin{aligned}
x(t+1)&=\begin{bsmallmatrix}
-1&0&-1\\-1.5&2.4&-4\\0&1&-1
\end{bsmallmatrix}x(t);\\
y(t)&=\begin{bsmallmatrix}
0.1&0.2&0.5
\end{bsmallmatrix}x(t),\hspace{5mm} x(0)\in X_{0}.
\end{aligned}\right.
\end{equation}
We can see that $\Sigma\times \Sigma_c$ is similar to $\Sigma_a\times \Sigma_{c_a}$, and once $(x_{a0},x_0)\in \mathcal{R}$, they will have the same output behavior.

In fact, if $\Sigma$ and $\Sigma_a$ are considered to be given beforehand with a simulation relation $\mathcal{R}$ from $\Sigma_a$ to $\Sigma$, in addition $\forall x_0\in X_0, \exists x_{a0}\in X_{a0}\mbox{ s.t. } (x_{a0},x_0)\in \mathcal{R}$. According to Algorithm 1, we can derive the related concrete and abstract DV systems $\Sigma_{\mbox{\scriptsize{DV}}}, \Sigma_{\mbox{\scriptsize{DV}}_a}$ such that $\Sigma_{\mbox{\scriptsize{DV}}}\cong \Sigma$ and $\Sigma_{\mbox{\scriptsize{DV}}_a}\cong\Sigma_a$, respectively. Afterwards, based on the transitivity of relations and initialization conditions, we can conclude that there exists a simulation relation $\mathcal{R}^d$ from $\Sigma_{\mbox{\scriptsize{DV}}_a}$ to $\Sigma_{\mbox{\scriptsize{DV}}}$ and in addition, $\forall x_0\in X_0, \exists x_{a0}\in X_{a0}\mbox{ s.t. } (x_{a0},x_0)\in \mathcal{R}$. Finally, according to Theorem \ref{solution1}, for any $\Sigma_{c_a}\in\mathfrak{C}(\Sigma_a)$, we can always refine it to $\Sigma_c$ such that $\Sigma_{c}\in\mathfrak{C}(\Sigma)$ and $\mathfrak{B}^\mathbf{y}_{\Sigma\times \Sigma_{c}}\subseteq \mathfrak{B}^\mathbf{y}_{\Sigma_a\times \Sigma_{c_a}}$.\QED
\end{exmp}

\section{Approximate control refinement for DAEs}
\noindent Since exact (bi)simulation relations cannot tolerate any error, there are obvious limitations for the system approximation that can be achieved. However, approximate relationships that do allow for the possibility of error, will certainly provide more freedom in control refinement. As a contrast of the exact control refinement, we will focus on approximate control refinement for DAE systems via the approximate simulation relations in this section. First of all, we introduce our previous research on hierarchical control for ODE systems, which immediately proposes an approach for developing approximate simulation relations and interfaces from the abstract models to the concrete models.
\subsection{Hierarchical control framework}
Consider a concrete ODE system
\begin{center}
$\Sigma_1:\left\{
\begin{aligned}
x(t+1)&=f(x(t),u(t));\\
y(t)&=g(x(t)),\hspace{5mm}x_0\in X_0,
\end{aligned}\right.$
\end{center}
where $x(t)\in X\subseteq \mathbb{R}^n,u(t)\in U\subseteq \mathbb{R}^p,y(t)\in Y\subseteq \mathbb{R}^k$. 

This discrete-time system is also a transition system $\Sigma_1=(X,U,X_{0},\rightarrow,Y,\mathcal{O})$ with:
\textit{
\begin{itemize}
\item the set of states is $X\subseteq \mathbb{R}^{n}$,
\item the set of inputs is $U\subseteq \mathbb{R}^p$,
\item the set of initial values is $X_{0}\subseteq X$,
\item the transition relation $\rightarrow:=(x,u,x^+)\in X\times U\times X\mbox{ s.t. } x^+ =f(x,u)$,
\item the set of outputs is $Y\subseteq \mathbb{R}^k$,
\item the output map is $\mathcal{O}:y=g(x)$.
\end{itemize}
}

An abstract ODE system $\Sigma_1^\prime$ that is developed via model reduction techniques is defined as 
\begin{center}
$\Sigma_1^\prime:\left\{
\begin{aligned}
z(t+1)&=h(z(t),v(t));\\
w(t)&=k(z(t)),\hspace{5mm}z_0\in Z_0,
\end{aligned}\right.$
\end{center}
with $z(t)\in Z\subseteq \mathbb{R}^m,v(t)\in V\subseteq \mathbb{R}^q,w(t)\in Y\subseteq \mathbb{R}^k$. Note that $\Sigma_1^\prime$ and $\Sigma_1$ have the same output space. We can derive in a similar way the corresponding transition system for $\Sigma_1^\prime$. 

According to Definition \ref{bridge}, a simulation function of $\Sigma_1^\prime$ by $\Sigma_1$ is a function over the Cartesian product of their state spaces explaining how a state trajectory of $\Sigma_1^\prime$ can be transformed into a state trajectory of $\Sigma_1$ such that the distance between the output behavior of the two systems remains bounded.

In the sequel, we will detail an approach for developing simulation functions, approximate simulation relations and interfaces for ODE systems. First of all, we introduce a special class of comparison functions, known as class $\mathcal{K}$ function~\cite{khalil1996nonlinear}. 
\begin{def1}
\textit{
A continuous function $\alpha:[0,a)\rightarrow[0,\infty)$ is said to belong to class $\mathcal{K}$ if it is strictly increasing and $\alpha(0)=0$. It is said to belong to class $\mathcal{K}_\infty$ if $a=\infty$ and $\alpha(r)\rightarrow \infty$ as $r\rightarrow \infty$.}
\end{def1}

One property of $\mathcal{K}$ function that will be used later is $\alpha \in \mathcal{K} \Rightarrow \alpha^{-1} \in \mathcal{K}$, where $\alpha^{-1}$ denotes the inverse function of $\alpha$.
\begin{lem}~\cite{jiang2001input}\label{lem1}
\textit{
For any $\mathcal{K}_\infty$ function $\alpha$ there is a $\mathcal{K}_\infty$ function $\hat{\alpha}$ satisfying\\
1. $\hat{\alpha}(s)\leq \alpha(s),\forall s\geq 0$;\\
2. $\eta-\hat{\alpha}\in \mathcal{K}$\\
where $\eta$ denotes the identity function or identity map, i.e., $\eta(x)=x$.} 
\end{lem}

Afterwards, let us detail the notion of simulation function and interface for a discrete-time system based on a Lyapunov-like auxiliary function and a level set. The construction here is different to that given in~\cite{main} for continuous-time systems. The idea of this Lyapunov-like auxiliary function comes from the theory of input-to-state stability~\cite{jiang2001input,khalil1996nonlinear}. Let us first construct a Lyapunov-like auxiliary function $\mathcal{V}:Z\times X\rightarrow \mathbb{R}^+$ together with a function $u_{\mathcal{V}}:V\times Z\times X\mapsto U$ such that for all $(z,x)\in Z\times X$,
\begin{equation}
\mathcal{V}(z,x)\geq \|k(z)-g(x)\| \label{eq14}
\end{equation} 
and for all $(v,z,x)\in V\times Z\times X$,
\begin{equation}
\begin{aligned}
\mathcal{V}(h(z,v),f(x,u_{\mathcal{V}}(v,z,x)))&-\mathcal{V}(z,x)\leq\\ &-\alpha(\mathcal{V}(z,x))+\sigma(\|v\|). \label{eq15}
\end{aligned}
\end{equation}
In (\ref{eq15}), $\alpha$ is a $\mathcal{K}_\infty$ function, $\sigma$ is a $\mathcal{K}$ function. Then we have the following proposition detailing the simulation functions for ODE systems.

\begin{prop}\label{constructsim}
\textit{Let $\mathcal{V}$ be a Lyapunov-like auxiliary function and $u_{\mathcal{V}}$ be a function such that (\ref{eq14}) and (\ref{eq15}) hold.
Then,
\begin{equation}\label{simulationfuction}
\mathcal{S}(z,x)=\max(\mathcal{V}(z,x),\gamma(v_{\max}))
\end{equation}
is a simulation function of $\Sigma_1^\prime$ by $\Sigma_1$ and $u_\mathcal{V}$ is an interface from $\Sigma_1^\prime$ to $\Sigma_1$. The constructed $\gamma$ function is given as
$$\gamma(r)=\hat{\alpha}^{-1}\left(\frac{\sigma(r)}{c}\right)$$ with $c\in (0,1]$. $\hat{\alpha}$ is the $\mathcal{K}_\infty$ function chosen according to Lemma \ref{lem1}. 
}
\end{prop}
\begin{rem}
Lyapunov-like auxiliary functions, simulation functions and interfaces are defined over vectors in some Euclidean spaces. As list in the Notation, $z,x$ represents time dependent signals. However, we use $z,x$ as vectors for these notions for simplicity in the expressions and proofs.  
\end{rem}

The proof of Proposition \ref{constructsim} is shown in Appendix \uppercase\expandafter{\romannumeral1}. Immediately, based on the properties of simulation functions, we obtain
\begin{equation}
\begin{aligned}
\|w(t)-y(t)\|&\leq S(z(0),x(0))\\&=\max(\mathcal{V}(z(0),x(0)),\gamma(v_{\max}))=\varepsilon. \label{eq16}
\end{aligned}
\end{equation}

Consequently, 
\begin{eqnarray}
\mathcal{R}_\varepsilon =\{(z,x)\in Z\times X\mid \mathcal{S}(z,x)\leq \varepsilon\} \nonumber
\end{eqnarray}
defines an approximate simulation relation from $\Sigma_1^\prime$ to $\Sigma_1$.

\subsection{Simulation functions for linear systems}
The application of the hierarchical control approach is based on computing a simulation function and the associated interface. In this subsection, we focus on a simple algorithm to construct simulation functions for linear standard state space systems. 

Consider the concrete and the abstract linear standard state space systems defined as
\begin{center}
$\Sigma_1:\left\{
\begin{aligned}
x(t+1)&=Ax(t)+Bu(t);\\
y(t)&=Cx(t),\hspace{5mm}x_0\in X_0,
\end{aligned}\right.$
\end{center}
where $x(t)\in X\subseteq \mathbb{R}^n,u(t)\in U\subseteq \mathbb{R}^p,y(t)\in Y\subseteq \mathbb{R}^k$ and
\begin{center}
$\Sigma_1^\prime:\left\{
\begin{aligned}
z(t+1)&=Fz(t)+Gv(t);\\
w(t)&=Hz(t),\hspace{5mm}z_0\in Z_0,
\end{aligned}\right.$
\end{center}
with $z(t)\in Z\subseteq \mathbb{R}^m,v(t)\in V\subseteq\mathbb{R}^q,w(t)\in Y\subseteq \mathbb{R}^k$. We assume, without loss of generality, that rank$(B)=p$, rank$(C)=k$ and $m\leq n$ since $\Sigma_1^\prime$ is simpler than $\Sigma_1$. Furthermore, we also assume that the concrete system $\Sigma_1$ is stabilizable. Thus, there exists a $p\times n$ matrix $K$ such that all the eigenvalues of matrix $A+BK$ are inside the unit disc in the complex plane. Whereafter, we have the following lemma for discrete-time cases and the lemma is developed referring to~\cite{main}, which deals with continuous-time cases.  
\begin{lem}~\cite{main} \label{main}
\textit{There exists a positive definite symmetric matrix M and a scalar number $\lambda\in (0,1)$ such that the following matrix inequalities hold:\begin{equation}
M\geq C^TC, \label{eq18}
\end{equation}
\begin{equation}
(A+BK)^TM(A+BK)\leq \lambda^2M. \label{eq19}
\end{equation}
}
\end{lem}

The computation method of the stabilizing $K$ and the positive definite symmetric matrix $M$ of (\ref{eq18}) and (\ref{eq19}) are shown in Appendix \uppercase\expandafter{\romannumeral2}, which is completely different from that of the continuous-time cases. 

We now give an approach to design the simulation function and the associated interface for a linear discrete-time system referring to the continuous-time cases in~\cite{main}. The proof of the following proposition is shown in Appendix \uppercase\expandafter{\romannumeral1} and is somehow different from that in~\cite{main}.
\begin{prop}\label{solvematrix}
\cite{main} 
\textit{Assuming that there exists an $n\times m$ matrix $P$ and a $p\times m$ matrix $Q$ such that the following linear matrix equations hold:
\begin{equation}
PF=AP+BQ, \label{eq20}
\end{equation}
\begin{equation}
H=CP. \label{eq21}
\end{equation}
Then, the function defined by$$\mathcal{V}(z,x)=\sqrt{(x-Pz)^TM(x-Pz)}$$ is a Lyapunov-like auxiliary function. Based on Proposition \ref{constructsim}, a simulation function of $\Sigma_1^\prime$ by $\Sigma_1$ is derived as
\begin{equation}\label{simulationfuction1}
\mathcal{S}(z,x)=\max(\mathcal{V}(z,x),\gamma(v_{\max})).
\end{equation}
The associated interface is given by 
\begin{equation}\label{interface}
u_\mathcal{V}(v,z,x)=Rv+Qz+K(x-Pz),
\end{equation}
where $v\in V, z\in Z, x\in X$, $R$ is an arbitrary $p\times q$ matrix.
}
\end{prop} 

In~\cite{main}, the author developed a similar proposition in order to construct the injective abstraction map $P$ and to attain the abstract system accordingly. But in our work, we employ model reduction methods to attain the abstract system firstly and then solve the matrix equations (\ref{eq20}) and (\ref{eq21}) to derive the projection matrix $P$ so as to establish connections between the concrete and abstract systems. As we can see, the linear matrix equations are the key ingredients to find the specific simulation function. We explore two approaches to solve the constrained Sylvester equations (\ref{eq20}) and (\ref{eq21}) via Kronecker product~\cite{laub2005matrix} and RQ factorization, respectively, see Appendix \uppercase\expandafter{\romannumeral2} for details.

\subsection{Main result: approximate control refinement for DAEs}
In this subsection, we focus on the solution of Problem 2. We still consider the concrete and abstract DAE systems defined as (\ref{cdaeee}) and (\ref{adaeeee}), respectively. We show that if there exists an approximate simulation relation $\mathcal{R}_\varepsilon$ from $\Sigma_a$ to $\Sigma$, in addition, $\forall x_0\in X_0, \exists x_{a0}\in X_{a0}\mbox{ s.t. } (x_{a0},x_0)\in \mathcal{R}_\varepsilon$,  then for any $\Sigma_{c_a}\in \mathfrak{C}(\Sigma_a)$, we can always refine $\Sigma_{c_a}$ to attain a controller $\Sigma_{c}$ for $\Sigma$ such that $\Sigma_{c}\in \mathfrak{C}(\Sigma)$ and $\mathfrak{B}^\mathbf{y}_{\Sigma_a\times \Sigma_{c_a}}\subseteq \mathcal{E}_\varepsilon\big(\mathfrak{B}^\mathbf{y}_{\Sigma\times \Sigma_{c}}\big)$.

Almost under the same settings of the exact control refinement as shown in the previous section. We still need the related DV systems $\Sigma_{\mbox{\scriptsize{DV}}}$ and $\Sigma_{\mbox{\scriptsize{DV}}_a}$ as (\ref{dv2new}) and (\ref{dv1new}) satisfying $\Sigma_{\mbox{\scriptsize{DV}}}\cong \Sigma$ and $\Sigma_{\mbox{\scriptsize{DV}}_a}\cong \Sigma_a$, respectively. As a consequence, using the transitivity of relations and the initialization conditions, we can also conclude that there exists an approximate simulation relation $\mathcal{R}_\varepsilon^d$ from $\Sigma_{\mbox{\scriptsize{DV}}_a}$ to $\Sigma_{\mbox{\scriptsize{DV}}_a}$, in addition $\forall x_0\in X_0, \exists x_{a0}\in X_{a0}\mbox{ s.t. } (x_{a0},x_0)\in \mathcal{R}_\varepsilon^d$. Thus, there exists a related interface $\mathcal{F}_\varepsilon:S_a\times X_a\times X\mapsto S$. 

According to Theorem \ref{bisimulationtheorem} and Theorem \ref{bisimulationtheorem1} and the interface $s=\mathcal{F}_\varepsilon(s_a,x_a,x)$, we develop the following theorem as a solution for Problem 2, which is similar to Theorem \ref{solution1}.
\begin{thm}\label{solution2}
\textit{Let $\Sigma$ and $\Sigma_a$ be the given concrete and abstract DAE systems defined as (\ref{cdaeee}) and (\ref{adaeeee}), respectively. $\mathcal{R}_\varepsilon$ is an approximate simulation relation from $\Sigma_a$ to $\Sigma$, and in addition $\forall x_0\in X_0, \exists x_{a0}\in X_{a0}\mbox{ s.t. } (x_{a0},x_0)\in \mathcal{R}_\varepsilon$. Then for any $\Sigma_{c_a}\in\mathfrak{C}(\Sigma_a)$ defined as (\ref{DAEcontroller}), the controller  
\begin{equation}\label{solution2law}
\Sigma_c:\left\{
\begin{aligned}
B_d^Tx(t+1)&=B_d^TA_dx(t)+B_d^TB_ds(t);\\
u(t)&=C_ux(t)+D_us(t),
\end{aligned}\right.
\end{equation}
with $s(t)=\mathcal{F}_\varepsilon(\mathcal{T}(x_a(t)),x_a(t),x(t))$, refines $\Sigma_{c_a}$ such that $\mathfrak{B}^\mathbf{y}_{\Sigma_a\times \Sigma_{c_a}}\subseteq \mathcal{E}_\varepsilon\big(\mathfrak{B}^\mathbf{y}_{\Sigma\times \Sigma_{c}}\big)$.}
\end{thm}
\vspace{1mm}

The proof of Theorem \ref{solution2} is also based on the proofs of Theorem \ref{bisimulationtheorem}, Theorem \ref{bisimulationtheorem1} and Lemma \ref{standarddaeexact}.  

On the other hand, for a given concrete DAE system $\Sigma$ with a related DV system $\Sigma_{\mbox{\scriptsize{DV}}}$, we can apply well-developed model reduction methods on $\Sigma_{\mbox{\scriptsize{DV}}}$ to attain an abstract DAE system $\Sigma_{\mbox{\scriptsize{DV}}_a}$, which can be rewritten into the related abstract DAE system $\Sigma_a$ via Algorithm 2. Since the matrix $A_d$ of $\Sigma_{\mbox{\scriptsize{DV}}}$ may have unstable eigenvalues. In these cases, we first use the stabilizing gain $K$ computed via Lemma \ref{main} to make $A_d+B_dK$ stable and then apply model reduction techniques. As presented in the previous subsection regarding the hierarchical control framework for standard state space systems, we can derive the approximate simulation relation $\mathcal{R}_\varepsilon^d$ together with the initialization conditions and the related interface $s=\mathcal{F}_\varepsilon(s_a,x_a,x)$. Finally, we can derive the approximate simulation relation $\mathcal{R}_\varepsilon$ from $\Sigma_a$ to $\Sigma$ together with the initialization conditions based on the transitivity of relations and initialization conditions.  
   
In the end of this section, we also consider a simple example as an interpretation.
\begin{exmp}\label{example3}
Consider the same given concrete DAE system $\Sigma=(E,A,B,C)$ as Example \ref{example2} with 
\begin{equation}\nonumber
E=\begin{bsmallmatrix}
1&0&0\\0&0&1\\0&0&0
\end{bsmallmatrix},A=\begin{bsmallmatrix}
-1&0&0\\0&1&0\\0&0&1
\end{bsmallmatrix},B=\begin{bsmallmatrix}
1\\1\\1
\end{bsmallmatrix},C=\begin{bsmallmatrix}
0.1\\0.2\\0.5
\end{bsmallmatrix}^T.
\end{equation}

$\Sigma_{\mbox{\scriptsize{DV}}}=(A_d,B_d,C_d,D_d,C_u,D_u)$ is the related DV system with
\begin{equation}\nonumber
A_d=\begin{bsmallmatrix}
-1&0&-1\\0&0&0\\0&1&-1
\end{bsmallmatrix},B_d=\begin{bsmallmatrix}
0\\-1\\0
\end{bsmallmatrix},C_u=\begin{bsmallmatrix}
0\\0\\-1
\end{bsmallmatrix}^T, D_u=0,
\end{equation}
and $C_d=\begin{bmatrix}
C_u^T&C^T
\end{bmatrix}^T$, $D_d=\begin{bmatrix}
D_u^T&0
\end{bmatrix}^T$. Then, the stabilizing $K=\begin{bmatrix}
0.1262&-0.8327&0.9843
\end{bmatrix}
$ is derived via Lemma \ref{main} and this results in a stable matrix $A_d+B_dK$. Afterwards, the two dimensional abstract DV system $\Sigma_{\mbox{\scriptsize{DV}}_a}=(A_{da},B_{da},C_{da},D_{da},C_{u_a},D_{u_a})$ is derived by applying balanced truncation model reduction technique to this stabilized system and 
\begin{equation}\nonumber
A_{da}=\begin{bsmallmatrix}
-0.051&0.123\\-0.123&-0.287
\end{bsmallmatrix},B_{da}=\begin{bsmallmatrix}
-1.683\\-1.675
\end{bsmallmatrix},C_a=\begin{bsmallmatrix}
0.889\\-0.747
\end{bsmallmatrix}^T,
\end{equation} 
\begin{equation}\nonumber
C_{ua}=\begin{bsmallmatrix}
-1.429\\1.499
\end{bsmallmatrix}^T, D_{ua}=0,C_{da}=\begin{bmatrix}
C_{ua}\\C_a
\end{bmatrix},D_{da}=\begin{bmatrix}
D_{ua}\\0
\end{bmatrix}.
\end{equation}

According to Algorithm 2, the abstract DAE system $\Sigma_a=(E_a,A_a,B_a,C_a)$ is developed and
\begin{equation}\nonumber
E_a=\begin{bsmallmatrix}
-0.705&0.709\\0&0
\end{bsmallmatrix},
A_a=\begin{bsmallmatrix}
-0.051&-0.29\\-1.429 &1.499
\end{bsmallmatrix},B_a=\begin{bsmallmatrix}
0\\-1
\end{bsmallmatrix}.
\end{equation}
 
According to Proposition \ref{solvematrix}, we can first design a Lyapunov-like auxiliary function $\mathcal{V}(x_a,x)$ together with the simulation function $\mathcal{S}(x_a,x)$. Afterwards, the approximate simulation relation from $\Sigma_{\mbox{\scriptsize{DV}}_a}$ to $\Sigma_{\mbox{\scriptsize{DV}}}$ is immediately defined as
\begin{equation}\nonumber
\mathcal{R}_\varepsilon^d=\{(x_a,x)\in X_a\times X\mid \mathcal{S}(x_a,x)\leq \varepsilon\} 
\end{equation}
with $\varepsilon=\mathcal{S}(x_{a0},x_0)$, and in addition $\forall x_0\in X_0, \exists x_{a0}\in X_{a0}\mbox{ s.t. } (x_{a0},x_0)\in \mathcal{R}_\varepsilon^d$. The related interface is 
\begin{equation}\label{interfacehaha1}
s=\mathcal{F}_\varepsilon(s_a,x_a,x)=Rs_a+Qx_a+K(x-Px_a),
\end{equation}
with $P, Q$ and $R$ solved via Proposition \ref{solvematrix} and 
\begin{equation}\nonumber
P=\begin{bsmallmatrix}
-1.1597&2.4387\\1.5254&-0.9658\\1.4005&-1.5960
\end{bsmallmatrix}, Q=\begin{bsmallmatrix}
-0.0410\\-0.4645
\end{bsmallmatrix}^T, R=\begin{smallmatrix}
0.955\end{smallmatrix}.
\end{equation}

Till here, we build up the framework as shown in Figure \ref{newstructure}. From the transitivity of relations and initialization conditions, we can conclude that $\mathcal{R}_\varepsilon=\{(x_a,x)\in X_a\times X\mid \mathcal{S}(x_a,x)\leq \varepsilon\}$ is an approximate simulation relation from $\Sigma_a$ to $\Sigma$, and in addition $\forall x_0\in X_0, \exists x_{a0}\in X_{a0}\mbox{ s.t. } (x_{a0},x_0)\in \mathcal{R}_\varepsilon$.

Now, let us consider a controller $\Sigma_{c_a}\in \mathfrak{C}(\Sigma_a)$ defined as
\begin{equation}\nonumber
\Sigma_{c_a}:\begin{bsmallmatrix}
1&1
\end{bsmallmatrix}x_a(t+1)=\begin{bsmallmatrix}
1&1
\end{bsmallmatrix}x_a(t)+u_a(t),
\end{equation}
and the closed loop is derived as
\begin{equation}\nonumber
\Sigma_a\times \Sigma_{c_a}:\left\{
\begin{aligned}
x_a(t+1)&=\begin{bsmallmatrix}
-0.179&1.458\\-0.25&1.04
\end{bsmallmatrix}x_a(t);\\
y_a(t)&=\begin{bsmallmatrix}
0.1&0.2&0.5
\end{bsmallmatrix}x_a(t),\hspace{5mm} x(a0)\in X_{a0},
\end{aligned}\right.
\end{equation}
 with $u_a(t)=\begin{bsmallmatrix}
-1.429&1.499
\end{bsmallmatrix}x_a(t)$. $\Sigma_a\times \Sigma_{c_a}$ is stable. Then according to Theorem \ref{bisimulationtheorem1}, we derive the control strategy for $\Sigma_{\mbox{\scriptsize{DV}}_a}$ as
\begin{equation}\nonumber
s_a(t):=\mathcal{T}(x_a(t))=\begin{bsmallmatrix}
0.076&-0.793
\end{bsmallmatrix}x_a(t).
\end{equation}
The controlled abstract DV system is the same as $\Sigma_a\times \Sigma_{c_a}$. Finally, according to Theorem \ref{solution1}, the refined controller $\Sigma_c$ for $\Sigma$ is derived as
\begin{equation}\nonumber
\Sigma_c: \left\{
\begin{aligned}
\begin{bsmallmatrix}
0&-1&0
\end{bsmallmatrix}x(t+1)&=Kx(t)+\begin{bsmallmatrix}
0.07&-0.763
\end{bsmallmatrix}x_a(t);\\
u(t)&=\begin{bsmallmatrix}
0&0&-1
\end{bsmallmatrix}x(t).
\end{aligned}\right.
\end{equation}

In the sequel, we choose the initial states $x_{a0}=\begin{bmatrix}
0.3&0.3
\end{bmatrix}^T
$ and $x_0=\begin{bmatrix}
0.4&0.2&-0.04
\end{bmatrix}^T$ such that $(x_{a0},x_0)\in \mathcal{R}_\varepsilon$. The simulation result of the closed loop systems is shown in Figure \ref{closed2}. Since the two controlled systems converge fast, we only show the simulation results of the closed loop systems until $t=15$.  
\begin{figure}[!h]
\centering
\includegraphics[width=1\columnwidth]{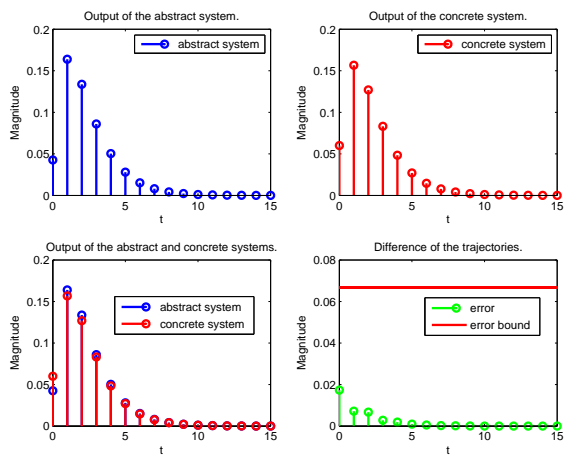}
\caption{Closed loop simulation results.}
\label{closed2}
\end{figure}

As we can see from Figure \ref{closed2}, the distance between the two controlled DAE systems is within the error bound 
\begin{equation}\nonumber
\varepsilon=\max(\mathcal{V}(x_{a0},x_0),\gamma({s_a}_{\max}))=0.0667.
\end{equation}
 
On the other hand, we consider the open loop simulation result by choosing a random signal $s_a$ to $\Sigma_{\mbox{\scriptsize{DV}}_a}$ satisfying ${s_a}_{\max}\leq 0.3$. The simulation result is shown in Figure \ref{simd3} with $\varepsilon=0.093$. 

\begin{figure}[!h]
\centering
\includegraphics[width=1\columnwidth]{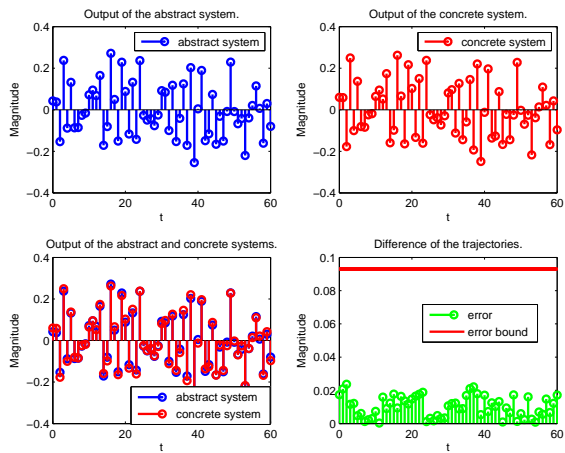}
\caption{Open loop simulation results.}
\label{simd3}
\end{figure}

It can be seen from Figure \ref{simd3}, the distance between the output behavior of the abstract and concrete DAE systems is bounded within $\varepsilon=0.093$.

Similar to the exact control refinement, if $\Sigma$ and $\Sigma_a$ are considered to be given beforehand with an approximate simulation relation $\mathcal{R}_\varepsilon$ from $\Sigma_a$ to $\Sigma$, in addition $\forall x_0\in X_0, \exists x_{a0}\in X_{a0}\mbox{ s.t. } (x_{a0},x_0)\in \mathcal{R}_\varepsilon$. According to Theorem \ref{solution2}, for any $\Sigma_{c_a}\in\mathfrak{C}(\Sigma_a)$, we can always refine it to $\Sigma_c$ such that $\Sigma_{c}\in\mathfrak{C}(\Sigma)$ and $\mathfrak{B}^\mathbf{y}_{\Sigma_a\times \Sigma_{c_a}}\subseteq \mathcal{E}_\varepsilon\big(\mathfrak{B}^\mathbf{y}_{\Sigma\times \Sigma_{c}}\big)$.\QED
\end{exmp}

\section{Conclusions}
\noindent In this paper, we dealt with the controller design problems of complex DAE systems that are frequently shown in industrial processes by developing control refinement approaches. These approaches were developed using the behavioral theory and the notions of (bi)simulation relations and approximate simulation relations from computer science. 

First of all, the behavioral approach was proposed as it introduces a general framework to treat dynamical systems. Afterwards, control problem and well-posed controllers for DAE systems were considered in the behavioral point of view. Then our control refinement problems were formulated in this behavioral framework. In order to acquire some insights, the properties of DAE systems were discussed and the related behavior of DAE systems was developed. In Section \uppercase\expandafter{\romannumeral3}, we also presented the notions of (bi)simulation relations and approximate simulation relations, which were widely mentioned in this paper. Followed by Section \uppercase\expandafter{\romannumeral4}, since it is difficult to deal with DAE systems directly, we introduced a calss of systems called driving variable systems that are behaviorally equivalent to the related DAE systems. Whereafter, two algorithms were developed for conversions between the DAE systems and the related DV systems. We also proved that a DAE system and its related DV system has a stronger relationship of bisimilarity. Subsequently, we concluded our control refinement framework for DAE systems, which illustrates the connections between DAE systems and their related DV systems and the connections between the concrete models and the abstract models. These connections are generated via (bi)simulation relations and approximate simulation relations.       

Based on the simulation relations and the initialization conditions between the abstract and the concrete DAE systems, we have proven that for any well-posed controller of the abstract DAE system, we can always refine it to attain a well-posed controller for the concrete DAE system such that they have the same controlled output behavior. As a contrast, approximate simulation relations that provide more freedom for controller design were considered. Whereafter, we proposed our approximate control refinement approach for DAE systems, which also introduces a new model reduction technique for DAE systems. In a similar way, on the basis of approximate simulation relations and the initialization conditions, we have proven that for any well-posed controller of the abstract DAE system, it can be refined to a well-posed controller for the concrete DAE system such that the distance between the output behavior of the two controlled systems is bounded within some error $\varepsilon$. 

The future research includes comparison of the control refinement approaches for DAE systems to results in perturbation theory and also control refinement for nonlinear DAE systems. On the other hand, the author is also interested in the application of geometric control theory in this topic.

\section*{Acknowledgements}
I would like to thank my supervisor Prof. Siep Weiland and my advisors Sofie Haesaert and Prof. Alessandro Abate. Their extremely helpful suggestions and invaluable assistance throughout this project during the past year are greatly appreciated. I would also like to thank the Control Systems Group and all my friends. This unforgettable year really means a lot to me. Finally, I want to thank my family for their support.

\bibliographystyle{abbrv}
\bibliography{mybib}{}

\begin{thebibliography}{10}

\bibitem{abadir2005matrix}
K.~M. Abadir and J.~R. Magnus.
\newblock {\em Matrix algebra}, volume~1.
\newblock Cambridge University Press, 2005.

\bibitem{antoulas2005approximation}
A.~C. Antoulas.
\newblock {\em Approximation of large-scale dynamical systems}, volume~6.
\newblock Siam, 2005.

\bibitem{Xingangthesis}
X.~Cao.
\newblock Hankel norm model reduction for discrete-time descriptor systems.
\newblock Master's thesis, Eindhoven University of Technology.

\bibitem{dai1989singular}
L.~Dai.
\newblock {\em Singular control systems}.
\newblock Springer-Verlag New York, Inc., 1989.

\bibitem{debeljkovic2011stability}
D.~Debeljkovic, I.~Buzurovic, and G.~Simeunovic.
\newblock Stability of linear discrete descriptor systems in the sense of
  lyapunov.
\newblock {\em International Journal of Information \& Systems Sciences}, 7(4),
  2011.

\bibitem{fainekos2007hierarchical}
G.~E. Fainekos, A.~Girard, and G.~J. Pappas.
\newblock Hierarchical synthesis of hybrid controllers from temporal logic
  specifications.
\newblock In {\em International Workshop on Hybrid Systems: Computation and
  Control}, pages 203--216. Springer, 2007.

\bibitem{linearconstrain}
A.~Girard and G.~J. Pappas.
\newblock Approximate bisimulations for constrained linear systems.
\newblock In {\em Decision and Control, 2005 and 2005 European Control
  Conference. CDC-ECC'05. 44th IEEE Conference on}, pages 4700--4705. IEEE,
  2005.

\bibitem{nonlinear}
A.~Girard and G.~J. Pappas.
\newblock Approximate bisimulations for nonlinear dynamical systems.
\newblock In {\em Decision and Control, 2005 and 2005 European Control
  Conference. CDC-ECC'05. 44th IEEE Conference on}, pages 684--689. IEEE, 2005.

\bibitem{metric}
A.~Girard and G.~J. Pappas.
\newblock Approximation metrics for discrete and continuous systems.
\newblock {\em Automatic Control, IEEE Transactions on}, 52(5):782--798, 2007.

\bibitem{main}
A.~Girard and G.~J. Pappas.
\newblock Hierarchical control system design using approximate simulation.
\newblock {\em Automatica}, 45(2):566--571, 2009.

\bibitem{bridge}
A.~Girard and G.~J. Pappas.
\newblock Approximate bisimulation: A bridge between computer science and
  control theory.
\newblock {\em European Journal of Control}, 17(5):568--578, 2011.

\bibitem{hespanha2009linear}
J.~P. Hespanha.
\newblock {\em Linear systems theory}.
\newblock Princeton university press, 2009.

\bibitem{jiang2001input}
Z.-P. Jiang and Y.~Wang.
\newblock Input-to-state stability for discrete-time nonlinear systems.
\newblock {\em Automatica}, 37(6):857--869, 2001.

\bibitem{khalil1996nonlinear}
H.~K. Khalil and J.~Grizzle.
\newblock {\em Nonlinear systems}, volume~3.
\newblock Prentice hall New Jersey, 1996.

\bibitem{kunkel2006differential}
P.~Kunkel and V.~L. Mehrmann.
\newblock {\em Differential-algebraic equations: analysis and numerical
  solution}.
\newblock European Mathematical Society, 2006.

\bibitem{laub2005matrix}
A.~J. Laub.
\newblock {\em Matrix analysis for scientists and engineers}.
\newblock Siam, 2005.

\bibitem{megawati2015bisimulation}
N.~Y. Megawati and A.~van~der Schaft.
\newblock Bisimulation equivalence of dae systems.
\newblock {\em arXiv preprint arXiv:1512.04689}, 2015.

\bibitem{moor2001robust}
T.~Moor and J.~M. Davoren.
\newblock Robust controller synthesis for hybrid systems using modal logic.
\newblock In {\em International Workshop on Hybrid Systems: Computation and
  Control}, pages 433--446. Springer, 2001.

\bibitem{mutsaers2012control}
M.~Mutsaers.
\newblock {\em Control relevant model reduction and controller synthesis for
  complex dynamical systems}.
\newblock PhD thesis, Technische Universiteit Eindhoven, 2012.

\bibitem{porru2015quality}
M.~Porru.
\newblock Quality regulation and energy saving through control and monitoring
  techniques for industrial multicomponent distillation columns.
\newblock 2015.

\bibitem{stykel2004gramian}
T.~Stykel.
\newblock Gramian-based model reduction for descriptor systems.
\newblock {\em Mathematics of Control, Signals and Systems}, 16(4):297--319,
  2004.

\bibitem{tabuada}
P.~Tabuada.
\newblock {\em Verification and control of hybrid systems: a symbolic
  approach}.
\newblock Springer Science \& Business Media, 2009.

\bibitem{van2004equivalence}
A.~Van~der Schaft.
\newblock Equivalence of dynamical systems by bisimulation.
\newblock {\em IEEE transactions on automatic control}, 49(12):2160--2172,
  2004.

\bibitem{weiland1991theory}
S.~Weiland.
\newblock {\em Theory of Approximation and Disturbnace Attenuation for Linear
  Systems}.
\newblock Rijksuniversiteit Groningen, 1991.

\bibitem{willems1998quadratic}
J.~Willems and H.~Trentelman.
\newblock On quadratic differential forms.
\newblock {\em SIAM Journal on Control and Optimization}, 36(5):1703--1749,
  1998.

\bibitem{willems1991paradigms}
J.~C. Willems.
\newblock Paradigms and puzzles in the theory of dynamical systems.
\newblock {\em Automatic Control, IEEE Transactions on}, 36(3):259--294, 1991.

\bibitem{willems2007behavioral}
J.~C. Willems.
\newblock The behavioral approach to open and interconnected systems.
\newblock {\em Control Systems, IEEE}, 27(6):46--99, 2007.

\bibitem{willems2013introduction}
J.~C. Willems and J.~W. Polderman.
\newblock {\em Introduction to mathematical systems theory: a behavioral
  approach}, volume~26.
\newblock Springer Science \& Business Media, 2013.

\end{thebibliography}

\clearpage
\begin{appendices}
\section{\textbf{Proofs}}
\subsection{Proof of Theorem \ref{pairwise}}
\begin{proof}
$\Sigma\cong \Sigma_{\mbox{\scriptsize{DV}}}$ can be proved based on the conditions of bisimulation relations and the initialization conditions. In this proof, we distinguish the states of $\Sigma$ and $\Sigma_{\mbox{\scriptsize{DV}}}$ by $x$ and $x_d$, respectively. Let us consider the relation $\mathcal{R}=\{(x,x_d)\in X\times X\mid x-x_d=0\}$, we first show that this $\mathcal{R}$ is a bisimulation relation between $\Sigma$ and $\Sigma_{\mbox{\scriptsize{DV}}}$. For any $(x,x_d)\in\mathcal{R}$, we have $Cx=Cx_d$ because they share the same output map. For any $(x,x_d)\in\mathcal{R}$ and any transition $x\xrightarrow{u} x^\prime$ in $\Sigma$, there exists transition $x_d\xrightarrow{s}_{\mbox{\tiny{DV}}} x_d^\prime$ in $\Sigma_{\mbox{\scriptsize{DV}}}$, with $s=W\begin{bmatrix}x^\prime&u&x\end{bmatrix}^T$ satisfying $(x^\prime,x_d^\prime)\in\mathcal{R}$, where $W$ is constructed via Algorithm 2. Conversely, for any transition $x_d\xrightarrow{s}_{\mbox{\tiny{DV}}} x_d^\prime$ in $\Sigma_{\mbox{\scriptsize{DV}}}$, there exists transition $x\xrightarrow{u} x^\prime$ in $\Sigma$ with $u=C_ux_d+D_us$ satisfying $(x^\prime,x_d^\prime)\in\mathcal{R}$, where $C_u$ and $D_u$ are constructed via Algorithm 1. Till here, we have proven that $\mathcal{R}$ is a bisimulation relation between $\Sigma$
and $\Sigma_{\mbox{\scriptsize{DV}}}$. In addition, For any $x_0\in X_0$, we can always find $x_{d0}=x_0\in X_0$ such that $(x_0,x_{d0})\in \mathcal{R}$ because they share the same initial state space. Similarly for any $x_{d0}\in X_0$, we can always find $x_{0}=x_{d0}\in X_0$ such that $(x_0,x_{d0})\in \mathcal{R}$. Finally, we prove that $\Sigma\cong \Sigma_{\mbox{\scriptsize{DV}}}$. 
\end{proof}

\subsection{Proof of Theorem \ref{bisimulationtheorem}}
\begin{proof}
We need to prove that this controller is well-posed and the two controlled systems are exactly the same. Since $u(t)=C_ux(t)+D_us(t)$, we obtain
\begin{equation}\nonumber
Ex(t+1)=(A+BC_u)x(t)+BD_us(t).
\end{equation}
Employing the control strategy (\ref{refinelaw}), the interconnected system is derived as 
\begin{equation}\nonumber
\begin{bmatrix}E\\B_d^T\end{bmatrix}x(t+1)=\begin{bmatrix}A+BC_u\\B_d^TA_d\end{bmatrix}x(t)+\begin{bmatrix}BD_u\\B_d^TB_d\end{bmatrix}s(t).
\end{equation}
Consider the the null space and right inverse of $\begin{bmatrix}E&-B\end{bmatrix}$, for which the computation details are shown in Appendix \uppercase\expandafter{\romannumeral2}, we obtain
\begin{equation}\nonumber
\begin{bmatrix}E&-B\end{bmatrix}\begin{bmatrix}B_d\\D_u\end{bmatrix}=0, \begin{bmatrix}E&-B\end{bmatrix}\begin{bmatrix}A_d\\C_u\end{bmatrix}=A,
\end{equation}
that is
\begin{equation}\nonumber
BD_u=EB_d, A+BC_u=EA_d.
\end{equation}
Therefore, the interconnected system is 
\begin{equation}\label{nonsquare}
\begin{bmatrix}E\\B_d^T\end{bmatrix}x(t+1)=\begin{bmatrix}E\\B_d^T\end{bmatrix}A_dx(t)+\begin{bmatrix}E\\B_d^T\end{bmatrix}B_ds(t).
\end{equation}
For two matrices $X\in \mathbb{R}^{m\times n}, Y\in \mathbb{R}^{n\times k}$, we know 
\begin{equation}\nonumber
rank(XY)\leq\min\{rank(X), rank(Y)\},
\end{equation}
and the Sylvester's rank inequality 
\begin{equation}\nonumber
rank(X)+rank(Y)\leq rank(XY)+n.
\end{equation}
If $X$ is full column rank, i.e., $rank(X)=n$, this will result in $rank(XY)=rank(Y).$\\
If $Y$ is full row rank, i.e., $rank(Y)=n$, this will result in $rank(XY)=rank(X).$

Therefore, in our case,
\begin{equation}\nonumber
rank\begin{pmatrix}
\begin{bmatrix}E_2\\B_d^T\end{bmatrix}\begin{bmatrix}I_n&B_d\end{bmatrix}
\end{pmatrix}=rank\begin{pmatrix}
\begin{bmatrix}E_2\\B_d^T\end{bmatrix}
\end{pmatrix}=n,
\end{equation}
\begin{equation}\nonumber
rank\begin{pmatrix}
\begin{bmatrix}E_2\\B_d^T\end{bmatrix}\begin{bmatrix}I_n&B_d&A_d\end{bmatrix}
\end{pmatrix}=rank\begin{pmatrix}
\begin{bmatrix}E_2\\B_d^T\end{bmatrix}
\end{pmatrix}=n.
\end{equation}
Since $\begin{bmatrix}E^T&B_d\end{bmatrix}^T$ is full column rank and has a left inverse, multiply the left inverse on both sides of (\ref{nonsquare}), the controlled system (\ref{nonsquare}) is transformed into 
\begin{equation}\nonumber
x(t+1)=A_dx(t)+B_ds(t),
\end{equation}
which is exactly the same as the driving variable system with the control strategy $s(t)$. Once the control strategy of $s(t)$ is determined, the controlled system has a unique trajectory. Hence, the refined controller is well-posed.
\end{proof}

\subsection{Proof of Theorem \ref{bisimulationtheorem1}}
\begin{proof} 
This proof is based on the computation details of Algorithm 2, which is shown in Appendix \uppercase\expandafter{\romannumeral2}. 

Consider the abstract DV system $\Sigma_{\mbox{\scriptsize{DV}}_a}$
\begin{equation} \label{haha1}
\Sigma_{\mbox{\scriptsize{DV}}_a}:\left\{
\begin{aligned}
x_a(t+1)&=A_{da}x_a(t)+B_{da}s_a(t);\\
u_a(t)&=C_{u_a}x_a(t)+D_{u_a}s_a(t);\\
y_a(t)&=C_ax_a(t)\hspace{0.5cm}x_a(0)\in X_{a0}.
\end{aligned}\right.
\end{equation} 
Where $x_a\in \mathbb{R}^m, s_a,u_a\in \mathbb{R}^q, y_a\in\mathbb{R}^k$.
First of all, according to Algorithm 2 for computation details) regarding the conversion from DV systems to DAE systems, we consider the SVD
\begin{equation}\nonumber
\mathcal{Q}=\begin{bmatrix}
A_{da}\\C_{u_a}
\end{bmatrix};
\mathcal{P}=\begin{bmatrix}
B_{da}\\D_{u_a}
\end{bmatrix}=U\Sigma V^T,
\end{equation}
where $U\in \mathbb{R}^{(m+q)\times(m+q)}, V\in \mathbb{R}^{q\times q}$ is unitary such that
$$UU^T=U^TU=I_{(m+q)\times(m+q)}, VV^T=V^TV=I_{q\times q}.$$
$$\Sigma=\left[\begin{matrix}
\bar{\Sigma}\\0_{m\times q}
\end{matrix}\right], \bar{\Sigma}=\left[\begin{matrix}
\sigma_1&0&0\\\vdots&\ddots&0\\0&\dots&\sigma_q
\end{matrix}\right], \sigma_1\geq\sigma_2\geq\dots\geq\sigma_q.$$
Similar to Algorithm 2, partition $U$ as
\begin{equation}\nonumber
U=\begin{bmatrix}
U_q&U_m
\end{bmatrix}=\begin{bmatrix}
U_3&U_1\\U_4&U_2
\end{bmatrix},
\end{equation}
where $U_q$ and $U_m$ represent the first $q$ and last $m$ columns of $U$ and 
$$U_q^TU_q=I_q, U_m^TU_m=I_m, U_m^TU_q=0_{m\times q}.$$
$U_3$ and $U_4$ represent the first $m$ and last $q$ rows of $U_q$, $U_1$ and $U_2$ represent the first $m$ and last $q$ rows of $U_m$. Whereafter, we obtain
\begin{equation}\nonumber
B_{da}=U_3\bar{\Sigma}V^T;
D_{u_a}=U_4\bar{\Sigma}V^T,
\end{equation}
or
\begin{equation}\label{hahaha}
B_{da}V{\bar{\Sigma}}^{-1}=U_3;
D_{u_a}V{\bar{\Sigma}}^{-1}=U_4.
\end{equation}

Now, consider the controlled system 
\begin{equation}\label{haha2}
\Sigma_a\times \Sigma_{c_a}:\left\{
\begin{aligned}
x_a(t+1)&=\mathcal{A}_ax_a(t);\\
y_a(t)&=C_ax_a(t),
\end{aligned}\right.
\end{equation}
with $u_a(t)=\mathcal{B}_ax_a(t)$. and the expression of 
\begin{equation}\label{haha3}
\begin{aligned}
s_a(t)&=V\bar{\Sigma}^{-1} \begin{bsmallmatrix}
U_q^T&-U_q^T\mathcal{Q}
\end{bsmallmatrix}\begin{bsmallmatrix}
x_a(t+1)^T&u_a(t)^T&x_a(t)^T
\end{bsmallmatrix}^T\\
&=V\bar{\Sigma}^{-1}\begin{psmallmatrix}U_3^Tx_a(t+1)+U_4^Tu_a(t)-U_q^T\mathcal{Q}x_a(t)\end{psmallmatrix}.
\end{aligned}
\end{equation}
Then, we substitute (\ref{haha2}) and (\ref{haha3}) to (\ref{haha1}) and use  (\ref{hahaha}) we obtain
\begin{equation}\label{haha5}
\begin{aligned}
x_a(t+1)&=A_{da}x_a(t)+U_3U_3^T(\mathcal{A}_ax_a(t))\\&+U_3U_4^Tu_a(t)-U_3U_q^T\mathcal{Q}x_a(t).
\end{aligned}
\end{equation}
\begin{equation}\label{haha6}
\begin{aligned}
u_a(t)&=A_{u_a}x_a(t)+U_4U_3^T(\mathcal{A}_ax_a(t))\\&+U_4U_4^Tu_a(t)-U_4U_q^T\mathcal{Q}x_a(t).
\end{aligned}
\end{equation}
According to (\ref{haha5}), once $u_a(t)=\mathcal{B}_ax_a(t)$ is chosen, $x_a(t+1)$ is uniquely determined based on $x_a(t)$ and the closed DV system is also autonomous. In order to simplify (\ref{haha5}) and (\ref{haha6}), we stack the two equations and obtain
\begin{equation}\label{haha7}
\begin{aligned}
\begin{bmatrix}x_a(t+1)\\u_a(t)\end{bmatrix}&=\mathcal{Q}x_a(t)+U_qU_3^T(\mathcal{A}_ax_a(t))\\&+U_qU_4^Tu_a(t)-U_qU_q^T\mathcal{Q}x_a(t).
\end{aligned}
\end{equation}
Multiply $U_q^T$ on both sides and use $U_q^TU_q=I_q$, we derive 
\begin{equation}\label{haha8}
\begin{aligned}
U_q^T\begin{bmatrix}x_a(t+1)\\u_a(t)\end{bmatrix}&=U_q^T\mathcal{Q}x_a(t)+U_3^T(\mathcal{A}_ax_a(t))\\&+U_4^Tu_a(t)-U_q^T\mathcal{Q}x_a(t).
\end{aligned}
\end{equation}
Since $U_q^T=\begin{bmatrix}
U_3^T&U_4^T
\end{bmatrix}$, finally, we obtain
\begin{equation}\nonumber
U_3^T(x_a(t+1)-\mathcal{A}_ax_a(t))=0,
\end{equation}
since $x_a(t+1)$ is solely based on $x_a(t)$ without any other freedom, we have 
\begin{equation}\nonumber
x_a(t+1)=\mathcal{A}_ax_a(t),
\end{equation}
Finally, we have proven that by applying the refined controller, the controlled DV system is the same as (\ref{haha2}).    
\end{proof}

\subsection{Proof of Proposition \ref{constructsim}}
\begin{proof}
Consider the Lyapunov-like auxiliary function $\mathcal{V}(z,x)$ satisfying (\ref{eq14}) and (\ref{eq15}). First, we denote $\mathcal{V}(h(z,v),f(x,u_{\mathcal{V}}(v,z,x)))$ by $\mathcal{V}^+(z,x)$ for convenience. Since $\hat{\alpha}$ is the $\mathcal{K}_\infty$ function chosen as Lemma \ref{lem1}, we have $\hat{\alpha}(s)\leq \alpha(s),\forall s\geq 0$. Therefore,
\begin{equation}
\mathcal{V}^+(z,x)-\mathcal{V}(z,x)\leq -\hat{\alpha}(\mathcal{V}(z,x))+\sigma(\|v\|). \label{eq17}
\end{equation}

For any input sequence $v$, consider the level set $$D=((z,x)|\mathcal{V}(z,x)\leq b)$$ where $b=\hat{\alpha}^{-1}\left(\frac{\sigma(v_{\max})}{c}\right)=\gamma(v_{\max})$. First we prove that when $(z(t_0),x(t_0))\in D$, $(z(t),x(t))\in D, \forall t\geq t_0$.

Assume that $(z(t_0),x(t_0))\in D$, $\mathcal{V}(z(t_0),x(t_0))\leq b$. With the inequality $\sigma(\|v\|)\leq \sigma(v_{\max})$, we transform (\ref{eq17}) into the following form:
\begin{equation}\nonumber
\begin{aligned}
\mathcal{V}^+(z(t_0),x(t_0))\leq -(1-c)&\hat{\alpha}(\mathcal{V}(z(t_0),x(t_0)))+ \\& \tilde{\alpha}(\mathcal{V}(z(t_0),x(t_0)))+\sigma(v_{\max}) 
\end{aligned}
\end{equation}
where $\tilde{\alpha}=\eta-c\hat{\alpha}$. Since $\eta-\hat{\alpha}\in \mathcal{K}$ as Lemma \ref{lem1}. In addition $\hat{\alpha}\in \mathcal{K}$, we have $(1-c)\hat{\alpha}\in \mathcal{K}$. Therefore, we can conclude that $\tilde{\alpha}=\eta-c\hat{\alpha}=\eta-\hat{\alpha}+(1-c)\hat{\alpha}\in \mathcal{K}$. 

Since $c\hat{\alpha}(\mathcal{V}(z(t_0),x(t_0)))\leq c\hat{\alpha}(b)=\sigma(v_{\max})$ and $c\tilde{\alpha}(\mathcal{V}(z(t_0),x(t_0)))\leq c\tilde{\alpha}(b)$, we have
\begin{equation}\nonumber
\begin{aligned}
\tilde{\alpha}(\mathcal{V}(z(t_0),x(t_0)))+\sigma(v_{\max})\leq &\tilde{\alpha}(b)+\sigma(v_{\max})=\\&b-c\hat{\alpha}(b)+\sigma(v_{\max})=b.
\end{aligned}
\end{equation}
Therefore,$$\mathcal{V}^+(z(t_0),x(t_0))\leq -(1-c)\hat{\alpha}(\mathcal{V}(z(t_0),x(t_0)))+b\leq b.$$
By induction, we can show that $(z(t_0+j),x(t_0+j))\in D,\forall j\in \mathbb{N}$, that is,$(z(t),x(t))\in D,\forall t\geq t_0$.\\
Now let $t_0=\min  \{t\in \mathbb{N}_0|(z(t),x(t)) \in D\}<\infty$.
Then$$\mathcal{V}(z(t),x(t))\leq \gamma(v_{\max}),\forall t\geq t_0.$$
For $0\leq t<t_0$, we have $c\hat{\alpha}(\mathcal{V}(z(t),x(t)))>c\hat{\alpha}(b)=\sigma(v_{\max})$. Therefore, $\forall 0\leq t<t_0$, we have
$$\mathcal{V}^+(z(t),x(t))-\mathcal{V}(z(t),x(t))\leq -(1-c)\hat{\alpha}(\mathcal{V}(z(t),x(t)))\leq 0.$$
We have proven that if $(z(0),x(0))\in D$, it will always remain in the level set and $(z(t),x(t))\in D, \forall t\in \mathbb{N}$. And if $(z(0),x(0))\notin D$, $\mathcal{V}(z(t),x(t))$ will decrease until $(z(t),x(t))$ gets in the level set and remains there.

Thus, by truncating the Lyapunov-like auxiliary function $\mathcal{V}(z,x)$ by the level set $\gamma(v_{\max})$, we construct the simulation function 
\begin{equation}\nonumber
\mathcal{S}(z,x)=\max(\mathcal{V}(z,x),\gamma(v_{\max}))
\end{equation}
such that (\ref{generalsim}) holds.
\end{proof}

\subsection{Proof of Proposition \ref{solvematrix}}
\begin{proof}
According to equation (\ref{eq18}) and (\ref{eq21}), we have $$\mathcal{V}(z,x)\geq \sqrt{(x-Pz)^TC^TC(x-Pz)}=\|Cx-Hz\|.$$Thus, inequality (\ref{eq14}) holds. To prove that inequality (\ref{eq15}) holds, we have
$$
\begin{aligned}
x^+-Pz^+ &= Ax+B[Rv+Qz+K(x-Pz)]\\&-P(Fz+Gv)\\&
= (A+BK)(x-Pz)+(BR-PG)v
\end{aligned}$$
where $x^+$ and $z^+$ denote the next states of $x$ and $z$ respectively. Therefore,\\

$
\mathcal{V}(h(z,v),f(x,u_{\mathcal{V}}(v,z,x)))-\mathcal{V}(z,x)=\\\\
\sqrt{(x^+-Pz^+)^TM(x^+-Pz^+)}
-\sqrt{(x-Pz)^TM(x-Pz)}\\\\
=\left \|\sqrt{M}[(A+BK)(x-Pz)+(BR-PG)v]\right \|-\\\\
\sqrt{(x-Pz)^TM(x-Pz)}.
$\\

\noindent From the triangle inequality of norms, we know that\\

$\mathcal{V}(h(z,v),f(x,u_{\mathcal{V}}(v,z,x)))-\mathcal{V}(z,x)\\\\
\leq \left \|\sqrt{M}(A+BK)(x-Pz)\right \|+\left \|\sqrt{M}(BR-PG)v\right \|-\sqrt{(x-Pz)^TM(x-Pz)}\\\\
=\sqrt{[(A+BK)(x-Pz)]^TM[(A+BK)(x-Pz)]}-\\\\
\sqrt{(x-Pz)^TM(x-Pz)}+\left \|\sqrt{M}(BR-PG)v\right \|.
$\\

\noindent Using inequality (\ref{eq19}), we obtain that\\

$\mathcal{V}(h(z,v),f(x,u_{\mathcal{V}}(v,z,x)))-\mathcal{V}(z,x)\\\\
\leq \lambda \sqrt{(x-Pz)^TM(x-Pz)}-\sqrt{(x-Pz)^TM(x-Pz)}+\left \|\sqrt{M}(BR-PG)v\right \|\\\\
=(\lambda-1)\mathcal{V}(z,x)+\left \|\sqrt{M}(BR-PG)v\right \|.
$\\

Hence, $\mathcal{V}(z,x)$ satisfies the two conditions (\ref{eq14}) and (\ref{eq15}) and it is a Lyapunov-like auxiliary function. In addition, $\gamma$ is the $\mathcal{K}$ function defined as
\begin{equation}
\gamma(r)=\frac{\left \|\sqrt{M}(BR-PG)\right \|_2}{1-\lambda}r. \label{eq22}
\end{equation}
Consequently, according to Proposition \ref{constructsim}, 
\begin{equation}\nonumber
\mathcal{S}(z,x)=\max(\mathcal{V}(z,x),\gamma(v_{\max}))
\end{equation}
is a simulation function and 
\begin{equation}\nonumber
u_\mathcal{V}(v,z,x)=Rv+Qz+K(x-Pz)
\end{equation}
is the associated interface.
\end{proof}

\subsection{Proof of Lemma \ref{vital}}
\begin{proof}
Since the anti-causal subsystem (\ref{acausal}) is reachable, rank ($\mathcal{R}_\mu)=n_2$ always holds as discussed in Proposition \ref{conmatrix}. Furthermore, $(N,B_2)$ is reachable if and only if rank$\left[\begin{matrix}
N-\lambda I&B_2
\end{matrix}
\right]=n_2, \forall \lambda\in \mathbb{C}$, which is similar to the PBH condition of normal state space systems. To prove that, assume $\left[\begin{matrix}
N-\lambda I&B_2
\end{matrix}
\right]<n_2$ for $\lambda$. Then, there exists a vector $w\neq 0$ such that 
$$w^T\left[\begin{matrix}
N-\lambda I&B_2
\end{matrix}
\right]=0,
$$ or
\begin{center}
$w^TN=\lambda w^T$ and $w^TB_2=0$.
\end{center}
Then $$w^TN^k=\lambda^kw^T$$ and $$w^TN^kB_2=\lambda^kw^TB_2=0, \forall k\in \mathbb{N}.$$Hence,
$$w^T\left[\begin{matrix} B_2&NB_2& \cdots & N^{\mu-1}B_2 \end{matrix}\right]=0.$$
So the reachability matrix $\mathcal{R}_\mu$ of the anti-causal subsystem is not full rank $n_2$ and finally derive the contradiction. Thus, rank$\left[\begin{matrix}
N-\lambda I&B_2
\end{matrix}
\right]=n_2, \forall \lambda\in \mathbb{C}$. Choose $\lambda=0\in \mathbb{C}$ and finally
\begin{equation}\nonumber 
rank\left[\begin{matrix}
N&B_2
\end{matrix}
\right]=rank\left[\begin{matrix}
N&-B_2
\end{matrix}
\right]=n_2.
\end{equation}
\end{proof}

\section{\textbf{Computation details}}
\subsection{Algorithm 1: DAE to DV conversion}
In order to rewrite DAE systems as DV systems, we first introduce the following lemma based on Proposition \ref{conmatrix} and refer to~\cite{dai1989singular} for details. The proof is shown in the end of Appendix \uppercase\expandafter{\romannumeral1}.
\begin{lem} \label{vital}
\textit{If the anti-causal subsystem (\ref{acausal}) is controllable, then rank$\left[\begin{matrix}
N&B_2
\end{matrix}
\right]=n_2$.}
\end{lem}

Consequently, we can also conclude that $\left[\begin{matrix}
E&-B
\end{matrix}
\right]$ has full row rank~\cite{dai1989singular}. Consider the state evolution equation of the DAE system (\ref{cdaeee}), we reorganize the state evolution as 
\begin{equation} \label{key}
\left[\begin{matrix}
E&-B
\end{matrix}\right]\left[\begin{matrix}
x(t+1)\\u(t)
\end{matrix}\right]=Ax(t). 
\end{equation}

Under the reachability assumption of the DAE system, we know that 
rank$\left[\begin{matrix}
E&-B
\end{matrix}
\right]=n$. We denote $\left[\begin{matrix}
E&-B
\end{matrix}
\right]$ by $M\in \mathbb{R}^{n\times (n+p)}$. $M^+\in \mathbb{R}^{(n+p)\times n}$ represents the Moore-Penrose pseudoinverse of $M$. Since $M$ has full row rank, it has a right inverse and we have $MM^+=I_{n}$. Then we use $B_{\mathcal{N}}\in \mathbb{R}^{(n+p)\times p}$ to represent the null space or the kernel of $M$ and obviously rank$(B_{\mathcal{N}})=p$.

Multiply an identity matrix by the right side of equation (\ref{key}), we obtain
$$M\left[\begin{matrix}
x(t+1)\\u(t)
\end{matrix}\right]=MM^+Ax(t).$$ The general solution is then given as
\begin{equation} \label{combine}
\left[\begin{matrix}
x(t+1)\\u(t)
\end{matrix}\right]=M^+Ax(t)+B_{\mathcal{N}}s(t)\end{equation}where $s(t)\in \mathbb{R}^p$. We regard $s(t)$ as the new driving input for the driving variable system. Partition $M^+$ and $B_{\mathcal{N}}$ into the first $n$ and last $p$ rows with
\begin{equation}\nonumber
M^+=\begin{bmatrix}
M_x\\M_u
\end{bmatrix}, B_{\mathcal{N}}=\begin{bmatrix}
B_{\mathcal{N}}^x\\B_{\mathcal{N}}^u\end{bmatrix}
\end{equation}

We derive the state space representation for equation (\ref{combine}) as
\begin{equation}
x(t+1)=M_xAx(t)+B_{\mathcal{N}}^xs(t),
\end{equation}
together with
\begin{equation}
u(t)=M_uAx(t)+B_{\mathcal{N}}^us(t).
\end{equation}

Till here, we have rewritten the DAE system (\ref{cdaeee}) into a DV system. 

\subsection{Algorithm 2: DV to DAE conversion}
Consider the related DV system $\Sigma_{\mbox{\scriptsize{DV}}}$ of the DAE system $\Sigma$, which is presented as
\begin{equation} \label{dv2newhaha}
\Sigma_{\mbox{\scriptsize{DV}}}:\left\{
\begin{aligned}
x(t+1)&=A_dx(t)+B_ds(t);\\
u(t)&=C_ux(t)+D_us(t);\\
y(t)&=Cx(t),\hspace{5mm}x(0)\in X_0,
\end{aligned}\right.
\end{equation} 
with $x(t)\in X\subseteq \mathbb{R}^n, s(t)\in S \subseteq \mathbb{R}^p, u(t)\in U\subseteq \mathbb{R}^p, y(t)\in Y\subseteq \mathbb{R}^k$. We denote that
$$\left[\begin{matrix}
A_d\\C_u
\end{matrix}\right]=\mathcal{Q},\left[\begin{matrix}
B_d\\D_u
\end{matrix}\right]=\mathcal{P}\in \mathbb{R}^{(n+p)\times p}. 
$$ 
We know that $rank(\mathcal{P})=p$, the singular value decomposition (SVD) of $\mathcal{P}$ is
\begin{equation}\label{svd}
\mathcal{P}=U\Sigma V^T
\end{equation}
with $U\in \mathbb{R}^{(n+p)\times(n+p)}, V\in \mathbb{R}^{p\times p}$ unitary, that is
$$UU^T=U^TU=I_{(n+p)\times(n+p)}, VV^T=V^TV=I_{p\times p}.$$
$$\Sigma=\left[\begin{matrix}
\bar{\Sigma}\\0_{n\times p}
\end{matrix}\right], \bar{\Sigma}=\left[\begin{matrix}
\sigma_1&0&0\\\vdots&\ddots&0\\0&\dots&\sigma_p
\end{matrix}\right], \sigma_1\geq\sigma_2\geq\dots\geq\sigma_p.$$
Hence, we derive that
\begin{equation} \label{svd}
\left[\begin{matrix}
x(t+1)\\u(t)
\end{matrix}\right]=\mathcal{Q}x(t)+U\Sigma V^Ts(t).
\end{equation}
Partition $U$ as $U=\left[\begin{matrix}
U_p&U_n
\end{matrix}\right]$, where $U_p$ and $U_n$ represent the first $p$ and last $n$ columns of $U$ and 
$$U_p^TU_p=I_p, U_n^TU_n=I_n, U_n^TU_p=0_{n\times p}.$$
Multiply $U_n^T$ on both sides of (\ref{svd}), the second item on the right vanishes because 
$$U_n^T\left[\begin{matrix}
U_p&U_n
\end{matrix}\right]\left[\begin{matrix}
\bar{\Sigma}\\0_{n\times p}
\end{matrix}\right]=\left[\begin{matrix}
0_{n\times p}&I_n
\end{matrix}\right]\left[\begin{matrix}
\bar{\Sigma}\\0_{n\times p}
\end{matrix}\right]=0_{n\times p}.$$ 
So we derive that
$$U_n^T\left[\begin{matrix}
x(t+1)\\u(t)
\end{matrix}\right]=U_n^T\mathcal{Q}x(t).$$
Partition $U_n^T$ as $U_n^T=\left[\begin{matrix}
U_1^T&U_2^T\end{matrix}\right]$, where $U_1^T$ and $U_2^T$ represent the first $n$ and last $p$ columns of $U_n^T$ and we obtain
$$U_1^Tx(t+1)=U_n^T\mathcal{Q}x(t)-U_2^Tu(t).$$
Finally, we transform the DV system (\ref{dv2newhaha}) back into the DAE system
\begin{equation} \label{dae}
\Sigma:\left\{
\begin{aligned}
U_1^Tx(t+1)&=U_n^T\mathcal{Q}x(t)-U_2^Tu(t);\\
y(t)&=Cx(t), \hspace{0.5cm}x(0)\in X_0.
\end{aligned}\right.
\end{equation} 
On the other hand, multiply $U_p^T$ on both sides of (\ref{svd}), we obtain
\begin{equation} \label{Sut}
U_p^T\left[\begin{matrix}
x(t+1)\\u(t)
\end{matrix}\right]=U_p^T\mathcal{Q}x(t)+\bar{\Sigma}V^Ts(t).
\end{equation}
Then, we can solve $s(t)$ as 
\begin{equation} \label{st}
\begin{aligned}
s(t)&=V\bar{\Sigma}^{-1} \left[\begin{matrix}
U_p^T&-U_p^T\mathcal{Q}
\end{matrix}\right]\left[\begin{matrix}
x(t+1)^T&u(t)^T&x(t)^T
\end{matrix}\right]^T\\
&=W\left[\begin{matrix}
x(t+1)^T&u(t)^T&x(t)^T
\end{matrix}\right]^T.
\end{aligned}
\end{equation}
\subsection{Solving the stabilizing gain K and the positive definite symmetric matrix M}
The stabilizing $K$ and the positive definite symmetric matrix $M$ of (\ref{eq18}) and (\ref{eq19}) can be computed by solving the semidefinite programming problem. Indeed, denoting $\hat{M}=M^{-1},W=KM^{-1}$ and using Schur complements, (\ref{eq18}) and (\ref{eq19}) are equivalent to the following matrix inequalities

$$\left[\begin{matrix}\hat{M}&\hat{M}C^T\\C\hat{M}&I_k\end{matrix}\right]\geq 0,$$\\
$$\left[\begin{matrix}-\lambda^2 \hat{M}&\hat{M}A^T+W^TB^T\\A\hat{M}+BW&-\hat{M}\end{matrix}\right]\leq 0.$$\\
\indent We choose a fixed parameter $\lambda$ and solve the above LMIs problem. However the simulation results we obtained are not very good since the coefficient of $\gamma$ function as shown in equation (\ref{eq22}) is too large. Then, we consider a better approach to solve (\ref{eq18}) and (\ref{eq19}).\\ 
\indent As shown in equation (\ref{eq22}), we can see the choice of the parameter $\lambda$ and the solved positive symmetric matrix $M$ will influence the coefficient of the $\gamma$ function. We try to solve (\ref{eq18}) and (\ref{eq19}) as an optimization problem to get the smallest coefficient and then we can apply an input sequence $v$ with the largest supremum norm. Therefore, we implement line search over the parameter $\lambda$ and try to minimize the trace of $M$ in order to obtain the smallest coefficient.\\
We rewrite (\ref{eq18}) as
\begin{equation}
M=\triangle M+C^TC \label{eq23}
\end{equation}
where $\triangle M$ is a positive definite symmetric matrix. And with (\ref{eq23}), we rewrite (\ref{eq19}) and try to solve (\ref{eq19}) as a LQR problem
\begin{equation}
(A-BK)^T(\triangle M+C^TC)(A-BK)\leq \lambda^2(\triangle M+C^TC). \label{eq24}
\end{equation}
We want to solve matrix inequality (\ref{eq24}) together with line search over $\lambda$. For each search, we can obtain the optimized positive definite symmetric matrix $\triangle M$. We divide both sides of (\ref{eq24}) by $\lambda^2$ and derive
\begin{equation}
(A_d-B_dK)^T(\triangle M+C^TC)(A_d-B_dK)\leq \triangle M+C^TC \label{eqhaha}
\end{equation}
where $A_d=\frac{A}{\lambda},B_d=\frac{B}{\lambda}$, $K$ is the gain matrix solved by LQR. (\ref{eqhaha}) is equivalent to
$$(A_d-B_dK)^T\triangle M(A_d-B_dK)-\triangle M$$
$$+\left[\begin{matrix}I&-K\end{matrix}\right]\left[\begin{matrix}A_d^TC^TCA_d-C^TC&A_d^TC^TCB_d\\B_d^TC^TCA_d&B_d^TC^TCB_d\end{matrix}\right]\left[\begin{matrix}I\\-K\end{matrix}\right]\leq 0.$$
Let $$L=\left[\begin{matrix}A_d^TC^TCA_d-C^TC&A_d^TC^TCB_d\\B_d^TC^TCA_d&B_d^TC^TCB_d\end{matrix}\right].$$
We conclude our optimization problem as
\begin{center}
minimize: $\operatorname{trace}(X)$\\
subject to: $X\succ 0,X\succ L.$  
\end{center}
The solution $$X=\left[\begin{matrix}Q&S\\S^T&R\end{matrix}\right]$$
determines the $Q,R,S$ matrix of a LQR problem. By solving the LQR problem we obtain $\triangle M$ and $K$. Finally, we get the optimized solutions of (\ref{eq18}) and (\ref{eq19}) which result in the smallest coefficient of (\ref{eq22}).
\subsection{Solving constrained Sylvester equations by Kronecker product}
Let $A\in \mathbb{R}^{m\times n},B\in \mathbb{R}^{p\times q}$. Then the \textit{Kronecker product (or tensor product)} of $A$ and $B$ is defined as 
$$A\otimes B=\left[\begin{matrix}
a_{11}B&\ldots&a_{1n}B\\ \vdots&\ddots&\vdots\\ a_{m1}B&\cdots&a_{mn}B
\end{matrix}\right]\in \mathbb{R}^{mp\times nq}.$$ 

Consider the linear matrix equation $AX+XB=C$ and rewrite it in terms of the columns, we obtain$$Ax_i+Xb_i=c_i=Ax_i+\sum_{j=1}^m b_{ji}x_j.$$
These equations can be rewritten as follows
$$\left[\begin{matrix}
A+b_{11}I&b_{21}I&\cdots&b_{m1}I\\ b_{12}I&A+b_{22}I&\cdots&b_{m2}I\\ \vdots&\vdots&\ddots&\vdots\\ b_{1m}I&b_{2m}&\cdots&A+b_{mm}I
\end{matrix}\right]\left[\begin{matrix}x_1\\\vdots\\x_m\end{matrix}\right]=\left[\begin{matrix}c_1\\\vdots\\c_m\end{matrix}\right].$$
Let $c_i\in \mathbb{R}^n$ denotes the columns of $C\in \mathbb{R}^{n\times m}$ so that $C=\begin{bmatrix}c_1,\cdots,c_m\end{bmatrix}$. Then $vec(C)$ is defined by stacking the columns of C on top of one another, i.e. $vec(C)=\begin{bmatrix}c_1^T&\cdots&c_m^T\end{bmatrix}^T\in \mathbb{R}^{mn}$.\\
\indent Finally the Sylvester equation $AX+XB=C$ can be rewritten in the following form as shown in ~\cite{laub2005matrix}
$$[(I_m\otimes A)+(B^T\otimes I_n)]vec(X)=vec(C).$$

\begin{prop}~\cite{laub2005matrix}
\textit{
Let $A\in \mathbb{R}^{n\times n},B\in \mathbb{R}^{m\times m},C\in \mathbb{R}^{n\times m}$. Then the Sylvester equation 
$$AX+XB=C$$ has a unique solution if and only if $A$ and $-B$ have no eigenvalues in common.}
\end{prop}

According to Kronecker product, we start to solve matrix equations (\ref{eq20}) and (\ref{eq21}). First we write (\ref{eq21}) as
$$(I_m\otimes C)vec(P)=vec(H)$$
and 
\begin{equation}
vec(P)=(I_m\otimes C)^+vec(H)+[I-(I_m\otimes C)^+(I_m\otimes C)]vec(Y) \label{eq26}
\end{equation}
where $vec(Y)$ is a vector with suitable length that needs to be determined and ``+" represents pseudoinverse. Then, (\ref{eq21}) is rewritten as
\begin{equation}
[-I_m\otimes A+F^T\otimes I_n]vec(P)=vec(BQ)=(I_m\otimes B)vec(Q). \label{eq27}
\end{equation}
Substituting (\ref{eq26}) into (\ref{eq27}), we obtain
$$[-I_m\otimes A+F^T\otimes I_n][I-(I_m\otimes C)^+(I_m\otimes C)]vec(Y)$$ $$=(I_m\otimes B)vec(Q)-[-I_m\otimes A+F^T\otimes I_n](I_m\otimes C)^+vec(H).$$

For convenience, we denote the above equation as
$$\mathcal{A}y=\mathcal{B}q+\mathcal{C},$$
where $y$ and $q$ represents $vec(Y)$ and $vec(Q)$, respectively. $\mathcal{A}$ is a square matrix. For a nonsingular $\mathcal{A}$ matrix, it is easy to solve. However, most of the cases, $\mathcal{A}$ is singular. Then we consider the following form
$$\left[\begin{matrix}\mathcal{A}&-\mathcal{B}\end{matrix}\right]\left[\begin{matrix}y\\q\end{matrix}\right]=\mathcal{C}.$$ We denote $\left[\begin{matrix}\mathcal{A}&-\mathcal{B}\end{matrix}\right]$ by $\mathcal{M}$, we have
$$\left[\begin{matrix}y\\q\end{matrix}\right]=\mathcal{M}^+vec(C)+(I-\mathcal{M}^+\mathcal{M})vec(X).$$
\indent For an arbitrary $vec(X)$ with suitable length, we obtain $vec(Y)$ and $vec(Q)$, and with the determined $vec(Y)$, we obtain $vec(P)$. Reshape the vectors into matrices forms, we get the solutions of $P$ and $Q$ matrices to equations (\ref{eq20}) and (\ref{eq21}) finally.

\subsection{Solving constrained Sylvester equations by RQ factorization} 
Consider the constrained Sylvester problem (\ref{eq20}) and (\ref{eq21}), we factorize $C$ into its RQ factorization as
$$C=[\begin{matrix}R_1&0 \end{matrix}] \left[\begin{matrix}
Q_1\\Q_2
\end{matrix}\right]$$
where $R_1\in \mathbb{R}^{k\times k}$ is full rank and $W_1=\left[\begin{matrix}
Q_1^T&Q_2^T
\end{matrix}\right]^T$ is an orthogonal matrix with  $W_1W_1^T=W_1^TW_1=I_n$, $W_1$ is partitioned into its first $k$ rows and its remaining $n-k$ rows. Then we obtain $$H=CP=R_1Q_1P$$ and all the solutions of $P$ are denoted by
\begin{equation}
P=Q_2^TZ+Q_1^TR_1^{-1}H. \label{eq28}\end{equation} 
Where $Z\in \mathbb{R}^{(n-k)\times m}$, substituting this in the Sylvester equation (\ref{eq20}) and multiplying on the left by the nonsingular matrix $W_1$, we obtain
$$\left[\begin{matrix}
Q_1\\Q_2
\end{matrix}\right]\left[Q_2^TZ+Q_1^TR_1^{-1}H\right]F-\left[\begin{matrix}
Q_1\\Q_2
\end{matrix}\right]A\left[Q_2^TZ+Q_1^TR_1^{-1}H\right]$$
$$=\left[\begin{matrix}
Q_1\\Q_2
\end{matrix}\right]BQ.$$
This yields the following two equations
$$R_1^{-1}HF-Q_1AQ_2^TZ-Q_1AQ_1^TR_1^{-1}H=Q_1BQ,$$
$$ZF-Q_2AQ_2^TZ-Q_2AQ_1^TR_1^{-1}H=Q_2BQ.$$
Then, we define $A_{11}=Q_1AQ_1^T$, $A_{12}=Q_1AQ_2^T$, $A_{21}=Q_2AQ_1^T$, $A_{22}=Q_2AQ_2^T$ and $B_1=Q_1B$, $B_2=Q_2B$, we obtain
\begin{equation}
R_1^{-1}HF-A_{12}Z-A_{11}R_1^{-1}H=B_1Q, \label{eq29}
\end{equation}
\begin{equation}
ZF-A_{22}Z-A_{21}R_1^{-1}H=B_2Q. \label{eq30}
\end{equation} 
\indent To simplify the two equations, we factorize the $k\times p$ matrix $B_1$ into its RQ factorization as:
$$B_1=[\begin{matrix}R_2&0 \end{matrix}] \left[\begin{matrix}
Q_3\\Q_4
\end{matrix}\right]$$
where $R_2\in \mathbb{R}^{k\times k}$ is full rank and $W_2=\left[\begin{matrix}
Q_3^T&Q_4^T
\end{matrix}\right]^T$ is an orthogonal matrix with 
$W_2W_2^T=W_2^TW_2=I_p$. Now let 
$$\hat{Q}=\left[\begin{matrix}
\hat{Q_3}\\\hat{Q_4}
\end{matrix}\right]=\left[\begin{matrix}
Q_3\\Q_4
\end{matrix}\right]Q.
$$Where $\hat{Q_3}\in \mathbb{R}^{k\times m}$ and $\hat{Q_4}\in \mathbb{R}^{(p-k)\times m}$. From equation (\ref{eq29}), we have
\begin{equation}
\hat{Q_3}=R_2^{-1}(R_1^{-1}HF-A_{12}Z-A_{11}R_1^{-1}H). \label{eq31}
\end{equation}
Using equation (\ref{eq30}) and let $$B_2W_2^T=[\begin{matrix}E_1&E_2\end{matrix}],$$we obtain
$$ZF-A_{22}Z-A_{21}R_1^{-1}H=[\begin{matrix}E_1&E_2\end{matrix}]\left[\begin{matrix}
\hat{Q_3}\\\hat{Q_4}
\end{matrix}\right]\\=$$
$$E_1R_2^{-1}R_1^{-1}HF-E_1R_2^{-1}A_{12}Z-E_1R_2^{-1}A_{11}R_1^{-1}H+E_2\hat{Q_4}$$
or
$$ZF+(E_1R_2^{-1}A_{12}-A_{22})Z=$$
\begin{equation}
E_1R_2^{-1}R_1^{-1}HF+A_{21}R_1^{-1}H-E_1R_2^{-1}A_{11}R_1^{-1}H+E_2\hat{Q_4}. \label{eq32}
\end{equation}
Finally, we reduce the original constrained Sylvester problem (\ref{eq20}) and (\ref{eq21}) to an unconstrained one. For each choice of $\hat{Q_4}$, (\ref{eq32}) has a unique solution, as long as the matrices $F$ and $A_{22}-E_1R_2^{-1}A_{12}$ have no common eigenvalues. So the solutions for matrices $P$ and $Q$ are 
$$P=Q_2^TZ+Q_1^TR_1^{-1}H,$$
$$Q=W_2^T\hat{Q}=[\begin{matrix}Q_3^T&Q_4^T\end{matrix}]\left[\begin{matrix}
\hat{Q_3}\\\hat{Q_4}
\end{matrix}\right]\\.$$Where $\hat{Q_3}$ is computed by (\ref{eq31}).

\end{appendices}

\end{document}